\documentclass[10pt,journal,compsoc]{IEEEtran}

\ifCLASSOPTIONcompsoc
  \usepackage[nocompress]{cite}
\else
  \usepackage[sort&compress]{cite}
\fi
\ifCLASSINFOpdf
  \usepackage[pdftex]{graphicx}
\else
\fi
\usepackage[cmex10]{amsmath}
\usepackage{dsfont}

\usepackage{mdwmath}
\usepackage{mdwtab}

\usepackage{eqparbox}

\usepackage{stfloats}
\usepackage{url}
\newcommand\MYhyperrefoptions{bookmarks=true,bookmarksnumbered=true,
pdfpagemode={UseOutlines},plainpages=false,pdfpagelabels=true,
colorlinks=true,linkcolor={black},citecolor={black},urlcolor={black},
pdftitle={An Efficient, Automatic Approach to High Performance Heterogeneous Computing},
pdfsubject={Typesetting},
pdfauthor={Gordon Inggs},
pdfkeywords={Heterogeneous Computing,Partitioning,Multicore-CPUs,GPUs,FPGAs,run-time Modelling,MILP}}
\ifCLASSINFOpdf
\usepackage[\MYhyperrefoptions,pdftex]{hyperref}
\else
\usepackage[\MYhyperrefoptions,breaklinks=true,dvips]{hyperref}
\usepackage{breakurl}
\fi
\usepackage{todonotes}
\usepackage{multirow}
\usepackage{listings}
\usepackage{caption}
\usepackage{subcaption}

\begin{document}
%
\title{A Domain Specific Approach to\\ High Performance Heterogeneous Computing}
%
%
%
%

\author{Gordon~Inggs,~\IEEEmembership{Student~Member,~IEEE,}
        David~B.~Thomas,~\IEEEmembership{Member,~IEEE,}
        and~Wayne~Luk,~\IEEEmembership{Fellow,~IEEE}
\IEEEcompsocitemizethanks{\IEEEcompsocthanksitem G.E. Inggs and D.B. Thomas are with the Circuits and Systems Group in the Department of Electrical and Electronic Engineering at Imperial College London\protect\\
E-mail: \href{mailto:gordon.e.inggs@ieee.org}{gordon.e.inggs@ieee.org}
\IEEEcompsocthanksitem W. Luk is with the Custom Computing Group in Department of Computing at Imperial College London}
}
\IEEEtitleabstractindextext{%
\begin{abstract}
Users of heterogeneous computing systems face two problems: firstly, in understanding the trade-off relationships between the observable characteristics of their applications, such as latency and quality of the result, and secondly, how to exploit knowledge of these characteristics to allocate work to distributed computing platforms efficiently. A domain specific approach addresses both of these problems. By considering a subset of operations or functions, models of the observable characteristics or domain metrics may be formulated in advance, and populated at run-time for task instances. These metric models can then be used to express the allocation of work as a constrained integer program, which can be solved using heuristics, machine learning or Mixed Integer Linear Programming (MILP) frameworks.

These claims are illustrated using the example domain of derivatives pricing in computational finance, with the domain metrics of workload latency or \emph{makespan} and pricing accuracy. For a large, varied workload of 128 Black-Scholes and Heston model-based option pricing tasks, running upon a diverse array of 16 Multicore CPUs, GPUs and FPGAs platforms, predictions made by models of both the makespan and accuracy are generally within 10\% of the run-time performance. When these models are used as inputs to machine learning and MILP-based workload allocation approaches, a latency improvement of up to 24 and 270 times over the heuristic approach is seen.
\end{abstract}}


\maketitle

\IEEEdisplaynontitleabstractindextext

%
\IEEEpeerreviewmaketitle

\ifCLASSOPTIONcompsoc
\IEEEraisesectionheading{\section{Introduction}\label{sec:introduction}}
\else
\section{Introduction}
\label{sec:introduction}
\fi

%
%


\IEEEPARstart{T}{he} following vignette illustrates the research problem that we address in this paper:

\textit{
Julia is a financial analyst at the Bank of England that monitors counterparty risk between investment banks. She is highly qualified in statistics and financial economics, and relies heavily on computational finance techniques to evaluate the derivative contracts that exist between investment banks. However, beyond the specialised programming environment that she uses, she knows little about computing and often runs her calculations for days on her laptop.
}

\textit{
She learns that a cluster of heterogeneous computing systems could massively accelerate her computations. She manages to cobble one together using the Bank's spare servers and cloud-based resources. Through the use of an open source application framework, she is soon able to execute her problems upon all of the heterogeneous computing platforms. However, she has no idea about how long a problem is going to take on a given platform. Furthermore, she is also mystified as to the relationship between the statistical accuracy she requires and the time it takes to evaluate her problems. Unable to understand the relationships between the metrics she cares about, she finds that some workloads take even longer on the cluster than on her laptop's CPU!
}

\textit{
Julia clearly needs a tool to help her not only understand the resources at her disposal, but also how to use them efficiently.
}

\break

\subsection{Problem Statement}

Julia might be a fiction, but the problems she faces are a reality for the increasing number of high performance computing application programmers. They have two problems:
\begin{enumerate}
\item Understanding the relationships between the run-time characteristics of their application tasks on heterogeneous computing platforms.
\item Allocating tasks to the available platforms so as optimise these run-time characteristics.
\end{enumerate}

In this paper, we both describe and demonstrate practically an approach to high performance, heterogeneous computing that addresses these problems. Our approach is premised on only supporting a subset of operations across all heterogeneous platforms. Computational application domains provide a natural means to limit the operations supported without overly inhibiting programmers, and hence our approach is a domain specific one.

We use the empirical definition of application domains as used in programming research~\cite{SmallMatterProgramming,DSL_bib,Fowler_DSL}, i.e. an identifiable category of computational activities where a small number of computational operations account for all or a disproportionately high proportion of the computations performed. For example, within the domain of Linear Algebra, vector arithmetic is used disproportionately more often than other operations. Hence, by focusing on supporting these frequently-used operations, these application domains can be practically supported across heterogeneous platforms.


\subsection{Contributions}
In this paper, we make the following contributions:

(1) We introduce a domain specific approach for modelling the run-time characteristics or metrics of heterogeneous computing platforms.

(2) We demonstrate metric modelling in the application domain of computational finance derivatives pricing. Our practical evaluation encompasses a large, diverse workload of 128 computational finance tasks across a heterogeneous computing cluster of 16 CPU, GPU and FPGA platforms across three continents.

(3) We show how the allocation of tasks to platforms can be formulated as a constrained integer programming problem. We demonstrate how the allocation problem can be solved using three distinct approaches: heuristics, machine learning and Mixed Integer Linear Programming (MILP).

(4) We apply the three allocation approaches to both synthetic and real world heterogeneous task and platform data. We show that while heuristics provide acceptable results, machine learning and MILP can provide orders of magnitude more efficient task allocations.


\subsection{Proposed Methodology}
We demonstrate that domain specific abstractions provide a means for characterising computing platforms in a manner that is \emph{meaningful} in the context of that domain, and hence to the domain programmer.


Furthermore, we show how this domain specific characterisation allows for heterogeneous platforms to be evaluated in a coherent manner, allowing for an efficient allocation of work across these resources.

\begin{figure}
\centering
\includegraphics[width=0.85\columnwidth]{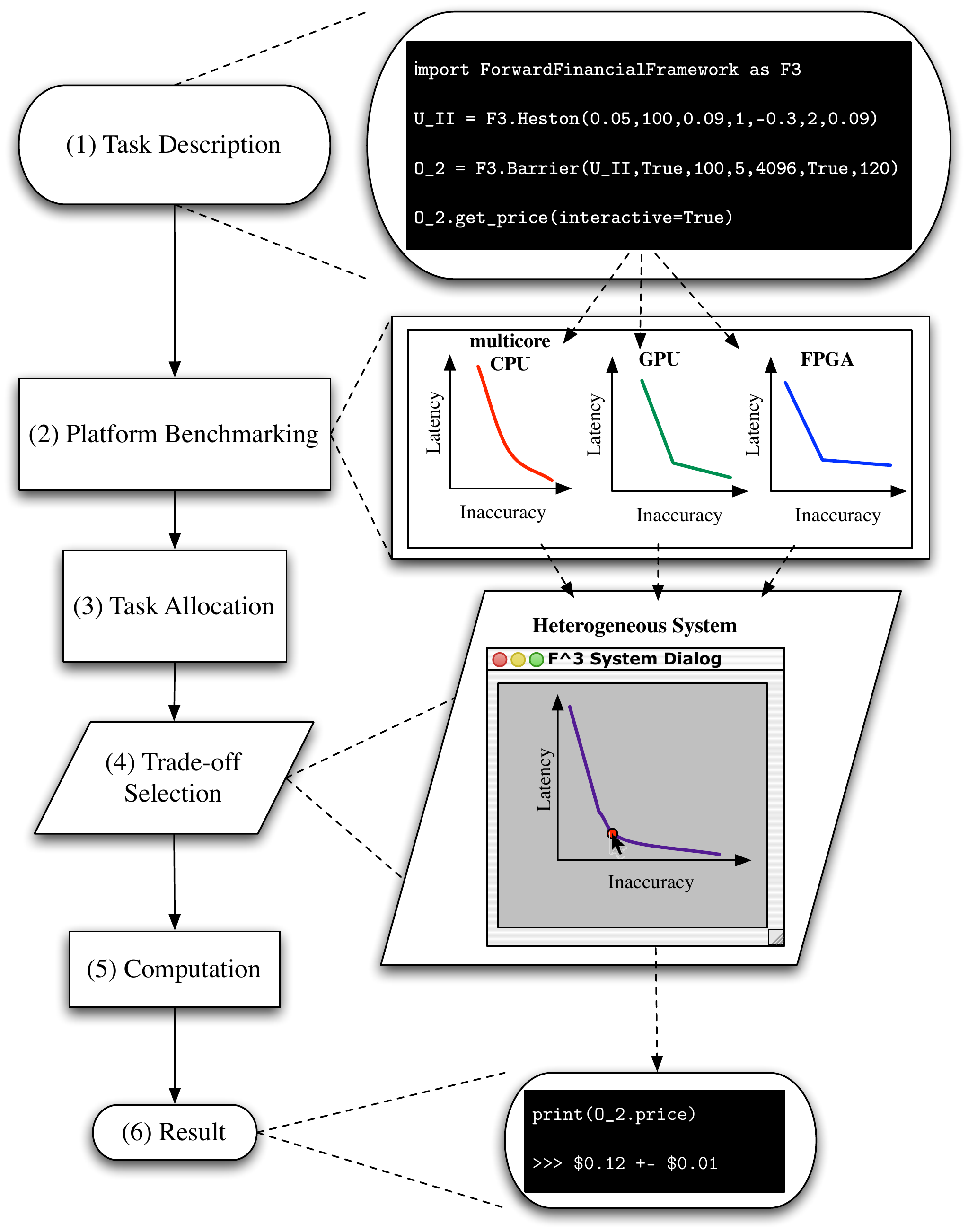}
\caption{Our proposed domain specific, high-level programming flow for high performance heterogeneous computing.}
\label{fig:F3InterfaceFlowchart}
\end{figure}

We seek to provide domain programmers such as Julia with the following programming flow, as illustrated in Figure \ref{fig:F3InterfaceFlowchart}:

(1) She specifies her tasks in a domain specific form. 

(2) Her tasks are then characterised using domain metrics with respect to the available platforms. 

(3) The optimal task allocations that make up the domain metric trade-off space are found automatically.

(4) Julia then selects the desired trade-off from the metric design space.

(5) Her workload is then evaluated, using the platforms in accordance with her objectives.

\subsection{The Rest of the Paper}
In Section \ref{sec:Background}, we elaborate on the background to the benefits of domain specific abstractions for heterogeneous computing, as well the state-of-the-art with respect to heterogeneous computing characterisation and workload allocation. We then expanded upon our two claims in Section \ref{sec:DomainCharacterisationPartitioning}: firstly, that domain specific abstractions enable the useful characterisation of heterogeneous platforms, and secondly, that these domain specific metric models can be used in partitioning work across heterogeneous platforms.


Then, in Section \ref{sec:CaseStudy} we demonstrate the domain specific methodology in practice by applying it to Julia's domain, financial derivatives pricing. We provide a brief overview of the domain and its heterogeneous implementation, after which we describe the latency and accuracy metric models, as well as heuristic, ML and MILP allocation approaches applied.  In Sections \ref{sec:DomainMetricModelEvaluation} and \ref{sec:DomainPartitionerEvaluation} we then evaluate our claims in the context of this case study.
 
 Finally we conclude the paper, summarising our major conclusions and lay out suggestions for further work.


\section{Background}
\label{sec:Background}

\subsection{Domain Specific Heterogeneous Computing}


An important finding in recent years is that domain specific abstractions can enable improved performance in a heterogeneous computing context~\cite{Delite,DSL_MonteCarlo,F3_P2S2}. As alluded to in the introduction, empirical studies of software engineering~\cite{SmallMatterProgramming} have found that a small set of algorithmic operations or design patterns within an application domain are executed disproportionately more frequently than others, often following a power law distribution. Indeed, application domains are often identified by grouping these operations together~\cite{DSL_bib}. By supporting the efficient, heterogeneous acceleration of these disproportionately influentially operations, significant gains can be realised automatically for programs restricted to a particular domain. 

Previous works have shown domain specific-enabled heterogeneous performance in practice, such as our own use of software application frameworks~\cite{F3_P2S2}, or domain specific languages, as shown by Chafi et al\cite{Delite} and Thomas and Luk\cite{DSL_MonteCarlo}. The key information yielded by the domain specific abstractions is the implicit dependency relationships between computations, allowing for heterogeneous parallelism to be exploited without programmer intervention.

However putting this approach into practice remains a challenge, requiring system developers with domain expertise to create domain specific abstractions~\cite{DSL_MonteCarlo,F3_P2S2} that support heterogeneous execution. Chafi et al's~\cite{Delite} approach advocates the use of language virtualisation, providing both a framework for creating implicitly parallel domain specific languages as well as a dynamic run-time for running applications created using such languages.



\subsection{Task Characterisation and Allocation \\on Heterogeneous Platforms}

The problem of characterising and allocating computational tasks to heterogeneous computing platforms has been widely studied for almost 40 years~\cite{HC_goals_methods_Open_Problems,HCChallengesOpportunities,Braun11Heuristics,TaskPartitioningMemoryConstrainedCPU,TaskAllocationDDP,AppLeS,SmartNet,StaticOpenCLScheduling,TaskAssignmentIteratedGreedy,HeuristicAlgorithmsSchedulingIndependentTasks, MILPHeterogeneous, FSP2015}. 


\subsubsection{Task Characterisation}

As identified by Braun et al~\cite{ HC_goals_methods_Open_Problems}, characterising the execution of tasks upon heterogeneous computing platform is comprised of three interrelated activities:

\emph{Task Profiling}: identifies the atomic (i.e. indivisible) tasks that comprises the current application. These tasks can then be further qualified by performing analysis or profiling of the task code. A key insight from Khokhar et al~\cite{HCChallengesOpportunities} is that profiling should determine the parallel execution modes possible for the given task. An increasingly popular approach is to require the programmer to identify the parallel execution modes, either through a specially designed API~\cite{Qilin} or by embedding this within the language itself~\cite{Delite}.

\emph{Analytic Platform Benchmarking}: identifies the capabilities of the heterogeneous computational platforms available. Another insight from Khockar et al~\cite{HCChallengesOpportunities} is that this process details how well the platform supports different parallel execution modes. A heterogeneous benchmark such as Rodinia~\cite{Rodinia} could be used for this purpose, or a representative subset of the current tasks.

\emph{Task-Platform Characterisation}: synthesises the data from the two previous activities, which results in models of how the specified tasks will execute upon the available resources. Grewe's work~\cite{StaticOpenCLScheduling} illustrates how a sophisticated machine learning-based approach can be used to do so.

As described in the next subsection, the last activity is usually not distinguished from allocating of tasks to platforms~\cite{StaticOpenCLScheduling,Qilin}. We argue that maintaining this separation is useful, as it allows for the quality of the characterisation activities to be evaluated independently from the allocation approach that is being used.

\subsubsection{The Allocation Problem}
\label{sec:PartitioningProblem}

When considering the allocation of tasks to heterogeneous computing resources, the general scenario considered in the literature, i.e.~\cite{Braun11Heuristics,TaskAllocationDDP,TaskAssignmentIteratedGreedy,HeuristicAlgorithmsSchedulingIndependentTasks,StaticOpenCLScheduling,TaskPartitioningMemoryConstrainedCPU, MILPHeterogeneous, FSP2015} is a set of independent or atomic tasks being partioned across or allocated to multiple heterogeneous platforms. It is assumed that a task will block a platform for its duration, i.e. occupy the computing resource completely. It is also commonly assumed that the allocation is being performed statically, in advance of the execution of any of the tasks.

In this general problem formulation, the general objective is to minimise the makespan. The makespan is the latency between when the first task is initiated until the last result returned for the task set. 


As described above, minimising the makespan with \emph{a priori} knowledge of the execution time of atomic tasks is a well studied problem. As we shall show in Section \ref{sec:DomainCharacterisationPartitioning}, as others have~\cite{MILPHeterogeneous}, this problem can be expressed formally as a 0-1 integer linear programming problem which has famously been shown to be NP-complete by Karp~\cite{Karp21Problems}.

\subsubsection{Allocation Approaches}
\label{sec:PartitionApproaches}
Surveying the literature, there are three categories of suggested approaches to the allocation problem described above:

\emph{Heuristic}~\cite{Braun11Heuristics,TaskAllocationDDP,HeuristicAlgorithmsSchedulingIndependentTasks}: a simple algorithmic rule is applied to allocate tasks to the available resources. Under specified circumstances such a rule might achieve a provably optimal allocation of tasks, and there is usually a lower bound on the quality of the solution relative to the optimal solution.

\emph{Machine Learning}~\cite{StaticOpenCLScheduling,TaskAssignmentIteratedGreedy}: a feasible task-platform allocation is improved using machine learning techniques such as Danzig's Simplex algorithm, simulated annealing or genetic algorithms. At worst these techniques can confirm the quality of the starting solution.

\emph{Integer Linear Programming}~\cite{TaskPartitioningMemoryConstrainedCPU,ILPPartitioningPipelinedScheduling,MILPHeterogeneous,FSP2015}: the problem can be formulated as a constrained integer program that can be solved using linear optimisation techniques. In addition to applying standard optimisation heuristics, a dual formulation of the problem can be used to prove the optimality of the solution.

\subsubsection{Analysis}

Generally heuristic approaches have been the most studied in the context of heterogeneous computing. Braun's comprehensive study~\cite{Braun11Heuristics} found that simpler heuristics achieve better results than more complex ones for the general case. This suggests that the truly optimal approach is case-specific, dependent upon the dynamics between the task and platforms concerned, and so the more complex an allocation approach, the more likely it is to map better to certain configurations than others.

ILP appears to be an understudied approach, usually applied only in environments of pressing resource constraint~\cite{TaskPartitioningMemoryConstrainedCPU}. This lack of attention is likely due to the NP-hard complexity of mixed integer linear programs and the NP-complete complexity of binary valued programs. 

However considerable progress has been made in ILP in the last three decades~\cite{WhyMILPWorks}, and hence we believe that this approach is now practical for run-time allocation~\cite{FSP2015}, as do others~\cite{MILPHeterogeneous}. A key insight is that an external measurement of solution quality is desirable so that a high quality solution that is not necessarily provably optimal can be identified.

\section{Domain Characterisation \& Allocation}
\label{sec:DomainCharacterisationPartitioning}

In this section we elaborate on our claims that a domain specific approach to heterogeneous computing allows for both the useful characterisation of task upon heterogeneous platforms, and in turn, an efficient allocation of those tasks to platforms. To illustrate our explanation, in this section we use examples from the domains of image filtering and linear algebra arithmetic. 

In Section \ref{sec:CaseStudy}, we apply our domain specific approach to the financial derivatives pricing domain.

\subsection{Characterising Tasks upon Platforms}

By useful characterisation, we mean actionable, i.e. the domain specific approach enables predictive modelling of the run-time characteristics of domain tasks upon a wide range of heterogeneous platforms. Characterisation would be useful to domain programmers such as Julia as it allows for the static comparison of different platforms which, as we will show in the next subsection, is critical in the efficient allocation of task workloads.

However, this domain specific characterisation is a contribution in its own right because it relates tasks and platforms using the fundamental concepts of the application domain. By modelling the task-platform relationship using domain metrics, the computational design space is made accessible to anyone working within that domain.

These models allow domain programmer to balance their objectives in terms they understand.


\subsubsection{Domain Metrics}


To find the computational design space for a task or group of tasks within an application domain, we first need to know what the dimensions of that design space should be, i.e. the quantitative measurements used within the domain. We define these quantitative characteristics of the domain \emph{metrics}.

While the actual metric used will vary from domain to domain, all fall into one of four categories:
\begin{itemize}
\item \emph{Latency}: the time between initiation and completion.
\item \emph{Throughput}: the rate at which the task is completed.
\item \emph{Quality}: the degree to a quantifiable goal is achieved.
\item \emph{Resource Use}: the resources used to complete the task.
\end{itemize}

For example, within the domain of image filtering, latency could be measured in the seconds required to filter an image, while throughput could be the number of images processed per second.

In the linear algebra arithmetic domain, quality might be measured as the unit of least precision in the calculations performed, while the resource use might be expressed using the average monetary cost per matrix arithmetic operation.

\subsubsection{Metric Models}

To predict metrics, we require models for how the task inputs map to the domain metrics on the target platform. We formalise these models in (\ref{eq:MetricMapping}): we seek model functions that map $p$ real-valued inputs to domain functions to $m$ real-valued metric values, i.e. $ \vec{F}: \vec{P} \rightarrow \vec{M} $, where $\vec{F}$ is the domain function model, $\vec{P}$ the inputs and $\vec{M}$ the metric values.

\begin{equation}
\label{eq:MetricMapping}
\begin{aligned}
 \vec{F}=(f_1,f_1,\cdots,f_{m}):  \vec{P} \rightarrow \vec{M} &  & \vec{P} \in \mathds{R}^p, \vec{M} \in \mathds{R}^m,\\
f_k(\vec{P}) = M_k &  & k = {1,2,\ldots,m}. \\
\end{aligned}
\end{equation}

As the application domain identifies in advance those operations which are disproportionately used, a function for mapping $\vec{P}$ to $\vec{M}$ of key domain functions for heterogeneous platforms can be found in advance.

For example, in the domain of image filtering, a candidate function for modelling would be the convolution operation used in all image filtering operations. In linear algebra arithmetic, the arithmetic operations would be modelled.


\subsubsection{Domain Variables and Parameters}

We refine the metric models further in (\ref{eq:ParametersDecomp}), $\vec{P}$ defines all possible input vectors to the domain specific operation. This space can be divided into two disjoint subsets, valid ($\vec{P_v}$) and invalid ($\vec{P_i}$) inputs. $\vec{P_i}$ are all of the inputs that will return a result that violates the correctness of the function as defined within the domain.

\begin{equation}
\label{eq:ParametersDecomp}
\begin{aligned}
& \vec{P} = \vec{P_i} \cup \vec{P_v} & \vec{P_i} \cap \vec{P_v} = \emptyset.
\end{aligned}
\end{equation}

For example, in the image filtering domain, when applying a uniform blur to an image, the set of inputs that define a non-uniformly weighted filter would be within $\vec{P_i}$ for that function. $\vec{P_v}$ is thus all of those inputs which return a valid result, representing the design space for that function.

By supplying the definition of ``correctness", the application domain makes explicit what input elements may be varied without affecting the correctness of the result. For example, in the linear algebra arithmetic domain, an input which specifies the maximum number of elements computed in parallel can be varied without affecting the correctness of the result.


We define those input elements which can be varied as \emph{domain variables} and those that cannot as \emph{domain parameters}. In our formalism, the domain definition identifies the subset of $\vec{P}$ upon which membership of $\vec{P_v}$ is defined.

\subsubsection{Identifying and Populating Metric Models}
The formalism above provides the criteria for potential metric model functions, however for each domain function there are an infinite number of possible metric model functions. When choosing one, we found that the simplest models to be the most broadly applicable.

For example, in the linear algebra arithmetic domain, a hypothetical metric model for latency of matrix-matrix addition operation might be expressed as the product of the size in the two matrices concerned ($N$) and the time per element-wise operation on that platform ($\alpha$), i.e. $$ f_L(N) = \alpha (N).$$

Similarly, in image filtering, the cost metric for applying a certain filter might be the cost per second of the platform ($\beta$) multiplied by the latency of the image processing ($f_L(S)$), i.e. $$ f_C(S) = \beta f_L(S).$$

As the structure of $\vec{F}$ is deterministic, an online benchmarking approach can be used to find the task and platform-specific metric model coefficients. We suggest a benchmarking procedure to generate a set of domain variable and metric values i.e. $\mathds{R}^{b \times p}$ and $\mathds{R}^{b \times m}$, where $b$ is the number of benchmarking iterations. The benchmarking data can then be used to to solve for the metric model function's coefficients.

We found weighted least squares regression to be effective in solving for the metric model coefficients. By using the variable benchmark values as weights, we reduced the impact of ``noise" present in metric measurements.

\subsection{Allocating Task to Platforms}
While the characterisation described in the previous subsection is useful when considering a heterogeneous platform in isolation, it is less helpful when faced with a cluster of heterogeneous platforms that can be used cooperatively. In this subsection we address how multiple computational domain metric model functions can be combined so as to create a unified, efficient design space.

\subsubsection{The General Allocation Problem}
We begin by expressing the makespan minimisation problem, as described in Section \ref{sec:PartitioningProblem}, as a binary valued integer linear program in (\ref{eq:MakespanMinimisation}).

Each non-zero element of the binary \emph{allocation} matrix ($\boldsymbol{A}$) represents an allocation of one of the $\tau$ tasks in a workload to one of the $\mu$ platforms, i.e. if $A_{i,j}=1$, then task $j$ has been allocated to platform $i$. The relative latency matrix ($\boldsymbol{L}$) gives the latency of each task upon each platform. Hence, similar to $A_{i,j}$, $L_{i,j}$ is the estimated relative latency of task $j$ upon platform $i$.


\begin{equation}
\label{eq:MakespanMinimisation}
\begin{aligned}
& \underset{\boldsymbol{A} \in \{0,1\}^{\mu \times \tau}}{\text{minimise}} & & G(\boldsymbol{A},\boldsymbol{L}) \quad \boldsymbol{L} \in \mathds R_+^{\mu \times \tau},\\
& \text{subject to} & & \sum\limits_{i=1}^\mu A_{i,j} = 1 \quad j =1,2,\ldots,\tau.\\
\end{aligned}
\end{equation}

where:
\begin{equation*}
\label{eq:MakespanMinimisationDefines}
\begin{aligned}
 & G(\boldsymbol{A},\boldsymbol{L}) = \max(\vec{H}(\boldsymbol{A},\boldsymbol{L})),\\
 & \vec{H}(\boldsymbol{A},\boldsymbol{L}) = (\boldsymbol{A} \circ \boldsymbol{L}) \cdot \boldsymbol{1}.\\
\end{aligned}
\end{equation*}

Reflecting the makespan minimisation problem's objective, we seek to minimise $G(\boldsymbol{A},\boldsymbol{L})$ while ensure that each task is completed, hence the constraint that the sum of each task entry, i.e. a column of $\boldsymbol{A}$, is 1.

This representation contains contains two reduction functions: firstly, the \emph{task latency reduction} ($\vec{H}(\boldsymbol{A},\boldsymbol{L})$), that is given by the element-wise multiplication or Hadamard product ($\boldsymbol{A} \circ \boldsymbol{L}$), dot multiplied by a vector of ones ($\boldsymbol{1}$); secondly, the \emph{platform latency reduction} ($G(\boldsymbol{A},\boldsymbol{L})$), that finds the maximum latency amongst the platforms for that allocation. 

These reduction functions map the allocation and task latency matrices ($\boldsymbol{A},\boldsymbol{L}$) to a vector of platform latencies, with an entry for each platform, and by which the vector of platform latencies are mapped to a scalar makespan value.

We now generalise this program, making use of the notion of domain metric models given in (\ref{eq:MetricMapping}). We assume that the valid variables $\vec{P}_v$ for each of the $\mu$ platforms are already known or can be easily approximated for each of the $\tau$ tasks. In (\ref{eq:AllocationGeneralised}) we seek an allocation ($\boldsymbol{A}$) so that we optimise the metric ($M_k$) for all tasks, as mapped by the task and platform reduction functions ($\boldsymbol{F}_k(\boldsymbol{A},\boldsymbol{P_v})$, $\vec{H}_k(\boldsymbol{A},\boldsymbol{P_v})$ and $G_k(\boldsymbol{A},\boldsymbol{P_v})$) into a scalar value.

\begin{equation}
\label{eq:AllocationGeneralised}
\begin{aligned}
& \underset{\boldsymbol{A}\in \{0,1\}^{\mu \times \tau}}{\text{optimise}} & & G_k(\boldsymbol{A},\boldsymbol{P_v}) \quad \boldsymbol{P_v} \in \mathds{R}^{\mu \times \tau \times p},\\
& \text{subject to} & & \sum\limits_{i=1}^\mu A_{i,j} = 1 \quad j =1,2,\ldots,\tau.\\
\end{aligned}
\end{equation}

where:
\begin{equation*}
\label{eq:AllocationGeneralisedDefines}
\begin{aligned}
 & G_k(\boldsymbol{A},\boldsymbol{P_v}): \vec{M_k} \to M_k \quad \vec{M_k} \in \mathds{R}^\mu, M_k \in \mathds{R},\\
 & \vec{H}_k(\boldsymbol{A},\boldsymbol{P_v}): \boldsymbol{M_k} \to \vec{M_k} \quad \boldsymbol{M_k} \in \mathds{R}^{\mu \times \tau},\\
 & \boldsymbol{F_k}(\boldsymbol{A},\boldsymbol{P_v}): (\boldsymbol{A},\boldsymbol{P_v}) \to \boldsymbol{M}_k.
\end{aligned}
\end{equation*}

\subsubsection{Splitting the Atomicity of Tasks}
Similar to the problem of heterogeneous characterisation, knowledge from the application domain can help find an efficient solution of the allocation problem. As structure of tasks is known in advance, the degree of parallelism within a task is known. As a result, allocation approaches can incorporate this information so as to allow for a task to be divided into subtasks while still providing a correct result. Making parallelism explicit enables a greater degree of work sharing between distributed computing resources, as discussed in Section \ref{sec:Background}. 

In this formulation, if the degree of parallelism is sufficiently large, this allows the elements of the allocation matrix, $\boldsymbol{A}$, to be ``relaxed", i.e. to be real-valued, and hence, the problem becomes linear and more tractable, as expressed in (\ref{eq:AllocationRelaxed}).

\begin{equation}
\label{eq:AllocationRelaxed}
\begin{aligned}
& \underset{\boldsymbol{A}\in \mathds R_+^{\mu \times \tau}}{\text{optimise}} & & G_k(\boldsymbol{A},\boldsymbol{P_v}) \quad \boldsymbol{P_v} \in \mathds{R}^{\mu \times \tau \times p},\\
& \text{subject to} & & \sum\limits_{i=1}^\mu A_{i,j} = 1 \quad j =1,2,\ldots,\tau.\\
\end{aligned}
\end{equation}

\subsubsection{Multimetric Pareto Surfaces}
As the metrics under consideration are also known ahead of execution, additional constraints may be added to the optimisation program for every other metric being considered ($M_x$), as described in (\ref{eq:AllocationConstrained}). This program requires that the allocation also satisfies all of the metric values specified in addition to optimising $M_k$.

\begin{equation}
\label{eq:AllocationConstrained}
\begin{aligned}
& \underset{\boldsymbol{A}\in \mathds R_+^{\mu \times \tau}}{\text{optimise}} & & G_k(\boldsymbol{A},\boldsymbol{P_v}) \quad \boldsymbol{P_v} \in \mathds{R}^{\mu \times \tau \times v},\\
& \text{subject to} & & \sum\limits_{i=1}^\mu A_{i,j} = 1 \quad j =1,2,\ldots,\tau,\\
& & & G_x(\boldsymbol{A},\boldsymbol{P_v}) = G_x \quad x \neq k,x = 1,2,\ldots,m.
\end{aligned}
\end{equation}

The multimetric optimisation program can be used to generate a Pareto surface, representing the heterogeneous computing platforms in terms of domain metric trade-offs. These trade-offs are achieved by changing the allocation of tasks to platforms.

For the metric Pareto surface to be populated, a range of values are required for all metrics that satisfy the program. This ranges of metrics can be found using the $\epsilon$-constraint method, as described by Kirlik and Say{\i}n~\cite{epsilonconstraint}.

This multimetric Pareto surface represents the culmination of our application of domain knowledge to heterogeneous computing. Our domain specific approach abstracts the allocation of task to platforms as the balancing of domain metrics. Hence, programmers such as Julia would be seamlessly able to use the capabilities of their heterogeneous platform by merely balancing their objectives.

\section{Case Study: Derivatives Pricing}
\label{sec:CaseStudy}
In this case study, we show how the domain specific modelling and allocation approach that we developed in the previous section can be applied to a new domain. We use the derivatives pricing domain, as we have done in other work~\cite{FSP2015}, in the broader area of computational finance, given its importance in global commerce.

We first describe the derivatives pricing domain, we then introduce the metric models of latency and accuracy that we use in our evaluation, and finally how the associated allocation problem can be solved using three different methods.

\subsection{Derivative Pricing Application Domain}
\label{sec:ComputationalFinanceDomain}

In this subsection we introduce the computational finance application domain, derivatives pricing, that we use as an example in the explanation of our approach and experiments to justify our claims. First, we describe derivatives pricing in general, and then define it as a computational domain. Finally, we describe an implementation of the domain. 


\subsubsection{Derivatives Pricing Background}

Computational finance is an important activity in modern commerce. The problems in the area are concerned with the quantitative measurement of uncertainty or risk. Derivatives pricing is one of the largest activities in this area, with \( \approx \) \$100 trillion of derivatives products currently active. Derivative pricing is also computationally intensive, and as a result is a major consumer of high performance computing, including multicore CPUs and GPUs.

An example of a derivative is an option contract. An option is contract where a holder pays a premium to the writer in order to obtain rights with regards to an underlying, an asset such as a stock or commodity. This right either allows the holder to buy or sell the underlying at a defined strike price at a defined exercise time. The holder has bought the \emph{right} to exercise the transaction if they so choose, and is in no way obligated to so. In derivatives pricing, the intrinsic value of the option is the payoff, the difference between the strike price and spot price of the underlying at the exercise time, or zero, whichever is higher~\cite{Hull}.

\begin{figure}
\centering
\includegraphics[width=.38\textwidth]{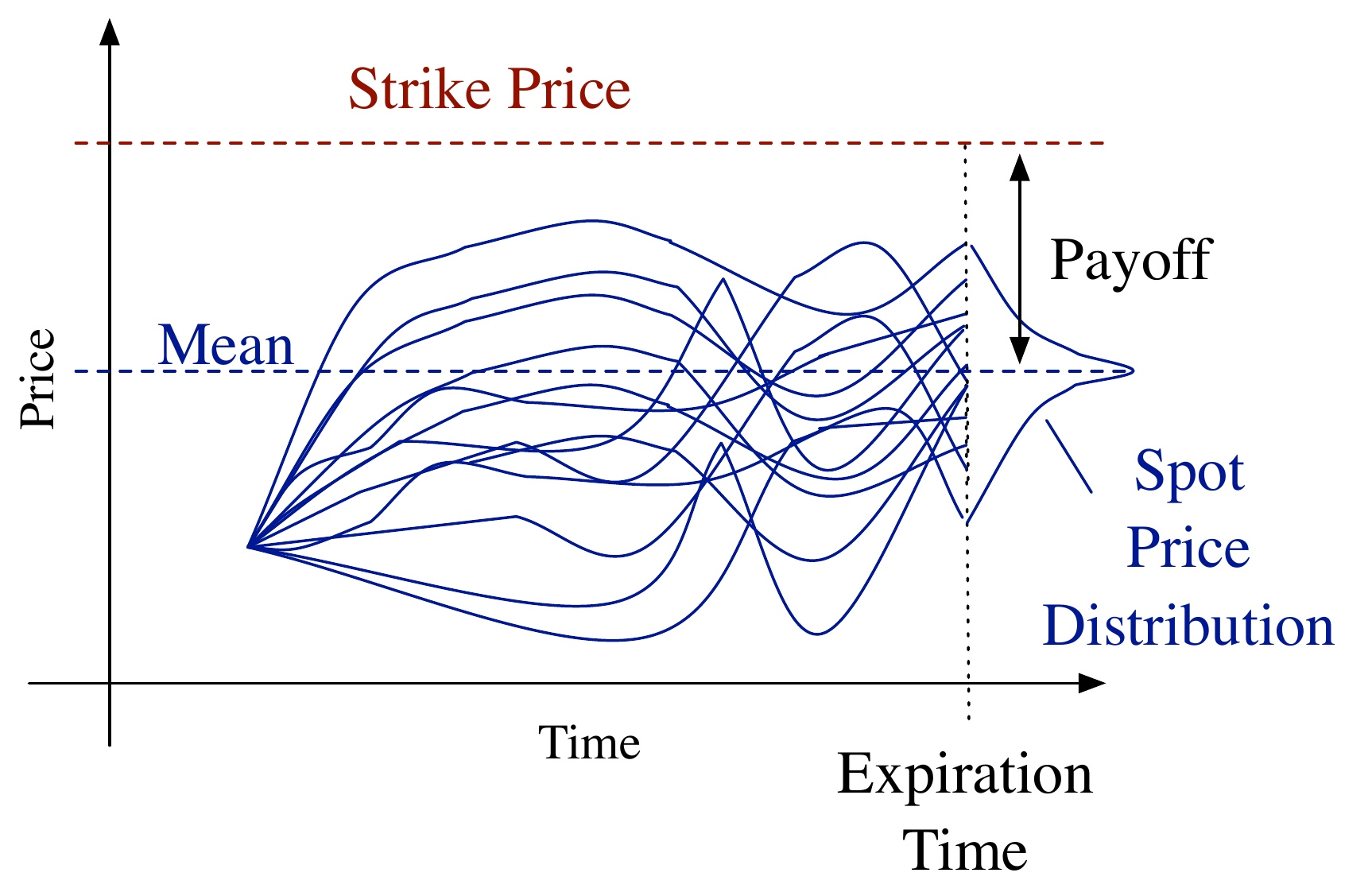}
\caption{Overview of Monte Carlo derivatives pricing.}
\label{fig:OptionMCOverview}
\end{figure}

The popular Monte Carlo technique for option pricing uses random numbers to create scenarios or simulation paths for the underlying based upon a model of its spot price evolution. The average outcome of these paths is then used to approximate the payoff~\cite{Hull}, as illustrated in Figure \ref{fig:OptionMCOverview}. Although computational expensive, this technique is robust, capable of tolerating underlying models with many more stochastic variables than competing methods~\cite{DSL_MonteCarlo,Hull}. Another advantage is that it is amenable to parallel execution. In fact, Monte Carlo is the canonical ``Embarrassingly Parallel'' algorithm~\cite{view_from_berkeley_2006}.

\subsubsection{Application Domain}
We now describe derivative pricing as an application domain in terms of types and functions.


\emph{The Underlying and Derivative Types}: the data in an option pricing task may be subdivided into two components, the derivative contract which is being priced and the underlying asset from which that derivative derives its value. The underlying encapsulates the probabilistic model, such as the Black-Scholes or Heston, being used to model the behaviour of the asset. The derivative embodies the details of the option contract both during the lifetime of the option as well at its expiration. 

The communication within a task can be formulated as a directed, acyclic graph, in which underlyings feed their prices to the derivatives which depend upon them.

\emph{Pricing Function}: The option pricing domain's sole function is finding the value of a type. Hence, the pricing function is typically only applied to the derivative type, as by definition an underlying type can provide its price at any point in time. Different techniques such as Monte Carlo or Tree-based methods could be used to implement the pricing function, provided the end result is the price of the derivative under consideration.





\subsubsection{Domain Implementation}
We now describe the Forward Financial Framework\footnotemark ($F^{3}$), an open source, financial domain application framework that we have developed, that implements the derivatives pricing domain.

\footnotetext{\url{https://github.com/Gordonei/ForwardFinancialFramework}}


\emph{Task description}: $F^3$ is implemented at the high level as a Python library~\cite{F3_P2S2}. A domain user, a financial engineer such as Julia, may use $F^{3}$'s classes to describe their derivatives pricing computations. There are three fundamental base classes that mirror the key concepts in the domain: derivatives, underlyings and solvers.

The underlying and derivative objects capture the attributes and behaviours of the underlying and derivative types as described above. The solver class is a collection for the derivatives that the programmer wishes to price as well as the platforms they wish to use.



\emph{Heterogeneous Implementations}: The solver class supports three behaviours upon heterogeneous platforms: code generation, compilation and execution. $F^3$ uses a wide array of back-end technologies: multicore CPUs using POSIX C; GPUs, Xeon Phi coprocessors and Altera FPGAs using the OpenCL standard~\cite{opencl}; Maxeler FPGAs using Maxeler's MaxJ.

All of the platform back-ends use a host-accelerator configuration, where a high performance coprocessor or subsystem is managed by a commodity CPU host. Communication between $F^3$ and platforms use the Secure Shell (SSH) protocol, allowing for tasks to be executed on remote platforms via TCP/IP networks.

\subsection{Financial Latency and Accuracy Metric Models}
As described in the previous subsection, for our example domain, pricing is the only function required. In this subsection, we develop the metric models, as per (\ref{eq:MetricMapping}), for the metrics of latency, (\ref{eq:LatencyModel}), and price accuracy, (\ref{eq:AccuracyModel}), for the pricing function as implemented using the Monte Carlo algorithm in $F^3$. This is in contrast to our other work, where we developed a financial cost metric model~\cite{FSP2015}.

\subsubsection{Latency Model} 

The latency between when a pricing operation is initiated and when it returns a price is fundamentally important within the financial domain~\cite{Hull}. This is because the time at which prices are received affects how traders use those prices. Minimising the latency of the pricing operation is desirable, as this confers first-mover advantage. 

We have used a simple, linear latency model in (\ref{eq:LatencyModel}), a function of a single domain variable, the number of simulation paths ($n$), i.e $n \in \mathds{Z},n \in \vec{P_v}$. The linear nature of the model reflects the $O(N)$ complexity of the Monte Carlo Algorithm. The model's coefficient ($\beta$) translates to the time spent per Monte Carlo path. Similarly, $\gamma$, the constant component of the latency metric model, captures the fixed time spent initialising the computation, as well as any time spent communicating the task to, and returning the result from, the target platform.

\begin{equation}
\label{eq:LatencyModel}
f_L(n) = \beta n + \gamma.
\end{equation}

\subsubsection{Accuracy Model}
In the financial domain, the accuracy of a computed price is expressed in probabilistic terms. When using the Monte Carlo technique, the size of 95\% confidence interval, as measured in currency of pricing (i.e. \$) is used. The accuracy measure is the size of the finite interval around the computed price for which there is a 95\% confidence that the true value lies within that interval. As small a confidence interval as possible is desired, as this means less variance in the price has to be accounted for.

The accuracy model that we used is based upon the convergence of the Monte Carlo algorithm, which is given by the inverse square root of the paths, scaled by a coefficient ($\alpha$). The model is given in (\ref{eq:AccuracyModel}).
\begin{equation}
\label{eq:AccuracyModel}
f_C(n) = \alpha n^{-\frac{1}{2}}.
\end{equation}

\subsubsection{Combined Model}

To relate the two domain metrics of latency and accuracy, we can solve for $n$ and use it to relate (\ref{eq:LatencyModel}) and (\ref{eq:AccuracyModel}) into a trade-off between the latency and accuracy ($c$), as given in (\ref{eq:CombinedModel}). 

\begin{equation}
\label{eq:CombinedModel}
\begin{aligned}
& & f_L(c) = \delta c^{-2} + \gamma.\\
\end{aligned}
\end{equation}

where:
\begin{equation*}
\label{eq:CombinedModelDefine}
\begin{aligned}
& \delta = \beta\alpha^2.\\
\end{aligned}
\end{equation*}

\subsection{Derivative Pricing Task Allocation}
We can now formulate the allocation problem using the derivative pricing metric models from the previous subsection, as well as outline three approaches for solving the problem.

\subsubsection{Reformulating the Allocation Problem}

In (\ref{eq:AppliedAllocation}) the unified domain metric model described in (\ref{eq:CombinedModel}) has been applied to the general, constrained allocation problem formulated in (\ref{eq:AllocationConstrained}). The vector $\vec{c}$ gives the required accuracies for the tasks, while $\boldsymbol{\gamma}$ is the task-platform constant matrix. Similarly, $\boldsymbol{\delta}:\vec{c}^2$ is the element-wise division of the delta coefficients by the required accuracies of the tasks. In this case, we have not to had to add an additional accuracy constraint, as the unified metric model has already captured this constraint.

\begin{equation}
\label{eq:AppliedAllocation}
\begin{aligned}
& \underset{\boldsymbol{A}\in \mathds R_+^{\mu \times \tau}}{\text{minimise}} & & G_L(\boldsymbol{A},\vec{c}) \quad \vec{c} \in \mathds{R}_+^\tau \\
& \text{subject to} & & \sum\limits_{i=1}^\mu A_{i,j} = 1 \quad j =1,2,\ldots,\tau.\\
\end{aligned}
\end{equation}

where:
\begin{equation*}
\label{eq:AppliedAllocationDefined}
\begin{aligned}
G_L(\boldsymbol{A},\vec{c}) &= \max(\vec{H_L}(\boldsymbol{A},\vec{c})),\\
\vec{H_L}(\boldsymbol{A},\vec{c}) &= (\boldsymbol{\delta}:\vec{c}^2 \circ \boldsymbol{A} +\boldsymbol{\gamma} \circ \lceil \boldsymbol{A} \rceil)\cdot \boldsymbol{1},\\
& \qquad \boldsymbol{\delta} \in \mathds R_+^{\mu \times \tau},\boldsymbol{\gamma} \in \mathds R_+^{\mu \times \tau} \\
\end{aligned}
\end{equation*}

An important feature of the formulation given in (\ref{eq:AppliedAllocation}) is its non-linearity as a result of the ceiling function in $\vec{H_L}(\boldsymbol{A},\vec{c}) $. This reflects (\ref{eq:LatencyModel}) and (\ref{eq:CombinedModel}), as there is a constant value for each platform-task entry, regardless of the scale of the allocation.



\subsubsection{Proportional Allocation Heuristic}



The first allocation approach, the proportional allocation heuristic, is given in (\ref{eq:ProportionalAllocation}). The heuristic allocates tasks inversely proportionally to the individual makespans of all of the platforms, attempting to balance tasks according to the relative capabilities of the different platforms.



\begin{equation}
\label{eq:ProportionalAllocation}
\vec{A}_{i,j} = \left (\vec{L}_i \sum^\mu_{o=1}{\frac{1}{\vec{L}_o}} \right ) ^{-1} \quad i = 1,2,\ldots,\mu,j = 1,2,\ldots,\tau.
\end{equation}

where:
\begin{equation*}
\label{eq:ProportionalAllocationDefines}
\begin{aligned}
& \vec{L} = \vec{H_L}(\boldsymbol{1},\vec{c}).\\
\end{aligned}
\end{equation*}

The heuristic only require an estimate of the relative latency of all tasks upon each platform. The proportional allocation heuristic works well provided the elements of $\boldsymbol{\gamma}$ are significantly smaller than the elements of $\boldsymbol{\delta}:\vec{c}^2$ for all platforms. If not, the tasks' cumulative constants dominate each platform's makespan, regardless of allocation. Theoretically, if there were no constant components, i.e. no setup time, then this heuristic would return the optimal allocation.


\subsubsection{Machine Learning Allocation}
The second approach uses the heuristic as a starting allocation of tasks. The platform reduction function $G_L(\boldsymbol{A},\vec{c})$ is then specified as the objective function for a time-constrained, global optimisation algorithm, the simulated annealing algorithm provided in SciPy~\cite{SCIPY}, combined with a ``polishing", convex optmisation algorithm, Danzig's Simplex algorithm, also available in SciPy. 


As this approach incorporates domain specific platform and task information as well as the heuristic, it should at worst confirm the heuristic, and at best find the most optimal allocation. As we will show in Section \ref{sec:DomainPartitionerEvaluation}, the objective function's linearity is a key determinate of the allocation optimality. Furthermore, another significant factor is problem size, as this problem suffers from the curse of dimensionality with respect to both $\mu$ and $\tau$.

\subsubsection{Mixed Integer Linear Programming Allocation}
The MILP approach uses the formulation of the domain allocation problem as the input to an open source, constrained integer programming framework, SCIP~\cite{SCIP}. SCIP applies global optimisation techniques as well as a variety of mathematical transformations and heuristics to solve the constrained problem.

SCIP accepts problems in a form very similar to (\ref{eq:AppliedAllocation}), expressed in Zuse Institut Mathematical Programming Language (ZIMPL)~\cite{ZIMPL}, which $F^3$ is capable of generating. However ZIMPL/SCIP does not allow for non-linear objective and constraint functions. This requires the problem to be reformulated as given in (\ref{eq:AppliedAllocationReformulated}), adding additional variables ($G_L$ and $\boldsymbol{B}$) and constraints to capture the non-linearities in the problem.

\begin{equation}
\label{eq:AppliedAllocationReformulated}
\begin{aligned}
& \underset{G_L,\boldsymbol{A}, \boldsymbol{B}} {\text{minimise}} & & G_L \quad G_L \in \mathds{R}_+,\boldsymbol{A}\in \mathds{R}_+^{\mu \times \tau}, \boldsymbol{B} \in \{0,1\}^{\mu \times \tau}, \\
& \text{subject to} & & \sum\limits_{i=1}^\mu A_{i,j} = 1 \quad j =1,2,\ldots,\tau, \\
& & & H_{L,i}(\boldsymbol{A},\vec{c}) \leq G_L \quad \vec{c} \in \mathds{R}_+^\tau, i = 1,2,\ldots,\mu, \\
& & & A_{i,j} \leq B_{i,j} \quad i = 1,2,\ldots,\mu,j = 1,2,\ldots,\tau.
\end{aligned}
\end{equation}
Where
\begin{equation*}
\label{eq:AppliedAllocationReformulatedDefined}
\begin{aligned}
\vec{H_L}(\boldsymbol{A},\vec{c}) &= (\boldsymbol{\delta}:\vec{c}^2 \circ \boldsymbol{A} +\boldsymbol{\gamma} \circ \boldsymbol{B} )\cdot \boldsymbol{1}.\\
\end{aligned}
\end{equation*}

Although binary integer linear programs are known to be NP-complete~\cite{Karp21Problems}, there has been progress in solving these problems efficiently~\cite{WhyMILPWorks}. 


\section{Evaluating Derivative Pricing Metrics}
\label{sec:DomainMetricModelEvaluation}
In this section, we evaluate our claim that the derivatives pricing metric models are able to characterise tasks on platform. 

To do so we need to evaluate the following two properties for the latency and accuracy models using a large, diverse set of platforms and tasks:
\emph{Incorporation}: When provided with additional information, the domain metric model predict the run-time value of that domain metric more accurately.
\emph{Extrapolation}: For a given amount of benchmarking, the domain metric values predicted by the models remains reasonably close to those seen at run-time for an increasing problem size.

To assess the degree to the properties were achieved, we measured the relative error ($E_k$) as given in (\ref{eq:RelativeError}), where the absolute difference between the predicted metric value ($f_k(n)$) and the run-time value ($\hat{f}_{k,n} $) is divided by the run-time value. The run-time metric value is measured when the task is run with the specified number of paths ($n$). 


\begin{equation}
\label{eq:RelativeError}
E_k = {{\left| f_k(n) - \hat{f}_{k,n} \right|}\over{\tilde{f}_{k,n}}}
\end{equation}

\subsection{Experimental Setup}

\subsubsection{Derivatives Pricing Tasks}
Table \ref{table:TaskDetails} provides an overview of the 128 option pricing tasks that were used to evaluate the financial domain metric models. In addition to the types of underlying and derivatives used, the total amount of computational work for each task is specified. 

The domain parameters for the pricing task operations, such as the proprieties of underlying model, were generated using uniform random numbers within the values of the Kaiserslautern option pricing benchmark~\cite{KS_Option_Benchmark}. We used a rejection procedure to keep the relative complexity of the pricing tasks within an order of magnitude.

\begin{table}

\caption{Evaluation workload of 128 derivative pricing tasks. Underlying types are Black-Scholes (\emph{BS}) and Heston (\emph{H}) model-based. Derivative types are Asian (\emph{A}), Barrier (\emph{B}), Double Barrier (\emph{DB}), Digital Double Barrier (\emph{DBB}) and European Options (\emph{E}).}
\label{table:TaskDetails}
\begin{tabular}{c|cccc}
    \parbox[c]{1.5cm}{Task Designation} & \parbox[c]{1cm}{Number} & \parbox[c]{1.2cm}{Underlying} & \parbox[c]{1cm}{Option} & \parbox[c]{1.5cm}{Computational Operations (kFLOP~/~path)}\\\hline 
    BS-A & 10 & BS & A & 139.267 \\ 
    BS-B & 10 & BS & B & 139.266 \\ 
    BS-DB & 10 & BS & DB & 143.360 \\ 
    BS-DDB & 5 & BS & DDB & 143.361 \\ 
    H-A & 25 & H & A & 319.492 \\ 
    H-B & 29 & H & B & 319.491 \\ 
    H-DB & 29 & H & DB & 323.585 \\ 
    H-DDB & 5 & H & DDB & 323.586 \\ 
    H-E & 5 & H & E & 315.395 \\ 
\end{tabular}

\end{table}

\subsubsection{Heterogeneous Platforms}
An overview of the heterogeneous platforms that we used are described in Table \ref{table:PlatformOverview}. The first class of platform heterogeneity is device type and manufacturer - we used a wide array of multicore CPUs, GPU and FPGA-based computational platforms from a variety of vendors.  The other is the diversity of interconnections used between the computational platforms, achieved with varied geographic locations.


The computational characteristics of the platforms are also described in Table \ref{table:PlatformOverview}. We describe the compute capabilities of the experimental platforms using the Kaiserslautern option pricing benchmark~\cite{KS_Option_Benchmark} and the Network Round-trip Time (RTT) as measured by the \emph{ping} network utility.

As the Monte Carlo algorithm being used is amenable to parallel execution, it is unsurprising that GPUs provide the best application performance, although an important caveat is that these performance figures are of implementations produced by $F^3$. A prominent data-point in terms of network latency is the Remote Server and Phi, which have orders of magnitude longer communication times than the other platforms due to being located in Cape Town, South Africa.

We expect the compute capabilities of platforms to determine the coefficient of the latency models ($\beta$) while the network latency will determine the constant coefficient, ($\gamma$).



\begin{table*}
\begin{center}

\caption{Overview of Experimental Heterogeneous Computing Platforms}
\label{table:PlatformOverview}
\centering
\begin{tabular}{c|cccccrr}
\parbox[l]{1.2cm}{Device Category} & \parbox[c]{1.5cm}{\center Device Designation} & \parbox[c]{1.2cm}{Device Vendor} & \parbox[c]{2.3cm}{Device Name} & \parbox[c]{1.5cm}{Network Location} & \parbox[c]{3cm}{Geographic Location} & \parbox[c]{1.8cm}{\center Application Performance (GFLOPS)} & \parbox[c]{1.5cm}{\center Network Round-trip Time (ms)} \\ \hline
 \multirow{9}{*}{\parbox[l]{1.2cm}{CPUs}}
    & Desktop &  Intel \textsuperscript{\textregistered} & \parbox[c]{2.3cm}{Core \textsuperscript{\textregistered} i7-2600} & Localhost & \parbox[c]{3.3cm}{ICL, London, UK} & 5.916 & 0.024 \\ 
    & Local Server & AMD \textsuperscript{\textregistered} & \parbox[c]{2.3cm}{Opteron \textsuperscript{\textregistered} 6272} & LAN & \parbox[c]{3.3cm}{ICL, London, UK} & 27.002 & 0.380 \\ 
    & Local Pi & ARM \textsuperscript{\textregistered} & \parbox[c]{2.3cm}{11 76JZF-S} & LAN & \parbox[c]{3.3cm}{ICL, London, UK} & 0.049 & 2.463\\ 
    & Remote Server & Intel \textsuperscript{\textregistered} & \parbox[c]{2.3cm}{Xeon \textsuperscript{\textregistered} E5-2680} & WAN & \parbox[c]{3.3cm}{UCT, Cape Town, ZA} & 11.523 & 3300.000 \\ 
    & AWS Server EC1 & Intel \textsuperscript{\textregistered} & \parbox[c]{2.3cm}{Xeon \textsuperscript{\textregistered} E5-2680} & WAN & \parbox[c]{3.3cm}{AWS, USA East Region} & 12.269 & 88.859 \\ 
    & AWS Server EC2 & Intel \textsuperscript{\textregistered} & \parbox[c]{2.3cm}{Xeon \textsuperscript{\textregistered} E5-2670} & WAN & \parbox[c]{3.3cm}{AWS, USA East Region} & 4.913 & 88.216 \\ 
    & AWS Server WC1 & Intel \textsuperscript{\textregistered} & \parbox[c]{2.3cm}{Xeon \textsuperscript{\textregistered} E5-2680} & WAN & \parbox[c]{3.3cm}{AWS, USA West Region} & 12.200 & 157.100\\ 
    & AWS Server WC2 & Intel \textsuperscript{\textregistered} & \parbox[c]{2.3cm}{Xeon \textsuperscript{\textregistered} E5-2670} & WAN & \parbox[c]{3.3cm}{AWS, USA West Region} & 4.926 & 159.578\\ 
    & GCE Server & Intel \textsuperscript{\textregistered} & \parbox[c]{2.3cm}{Xeon \textsuperscript{\textregistered}} & WAN & \parbox[c]{3.3cm}{GCE, USA Central Region} & 6.022 & 111.232\\    \hline
 \multirow{5}{*}{\parbox[l]{1.2cm}{GPUs}}
    & Local GPU 1 & AMD\textsuperscript{\textregistered} & \parbox[c]{2.3cm}{FirePro \textsuperscript{\textregistered} W5000} & LAN & \parbox[c]{3.3cm}{ICL, London, UK} & 212.798 & 0.269 \\ 
    & Local GPU 2 & Nvidia\textsuperscript{\textregistered} & \parbox[c]{2.3cm}{Quardo \textsuperscript{\textregistered} K4000} & LAN & \parbox[c]{3.3cm}{ICL, London, UK} & 250.027 & 0.278 \\ 
    & Remote Phi & Intel\textsuperscript{\textregistered} & \parbox[c]{2.3cm}{Xeon Phi \textsuperscript{\textregistered} 3120P} & WAN & \parbox[c]{3.3cm}{UCT, Cape Town, ZA} & 70.850 & 3300.000 \\ 
    & AWS GPU EC & Nvidia\textsuperscript{\textregistered} & \parbox[c]{2.3cm}{Grid \textsuperscript{\textregistered} GK104}  & WAN & \parbox[c]{3.3cm}{AWS, USA East Region} & 441.274 & 88.216 \\ 
    & AWS GPU WC & Nvidia\textsuperscript{\textregistered} & \parbox[c]{2.3cm}{Grid \textsuperscript{\textregistered} GK104} & WAN & \parbox[c]{3.3cm}{AWS, USA West Region} & 406.230 & 159.578 \\    \hline
 \multirow{2}{*}{\parbox[l]{1.2cm}{FPGAs}}
    & Local FPGA 1 & Xilinx\textsuperscript{\textregistered} & \parbox[c]{2.3cm}{Virtex \textsuperscript{\textregistered} 6 475T} & LAN & \parbox[c]{3.3cm}{ICL, London, UK} & 114.590 & 0.217 \\ 
    & Local FPGA 2 & Altera\textsuperscript{\textregistered} & \parbox[c]{2.3cm}{Stratix \textsuperscript{\textregistered} V D5} & LAN & \parbox[c]{3.3cm}{ICL, London, UK} & 161.074 & 0.299 \\ 
\end{tabular}

\end{center}
\end{table*}

\subsection{Model Error Results}
\subsubsection{Latency Model}
\begin{figure*}%
\centering

\begin{subfigure}{.75\columnwidth}
        \includegraphics[width=\columnwidth]{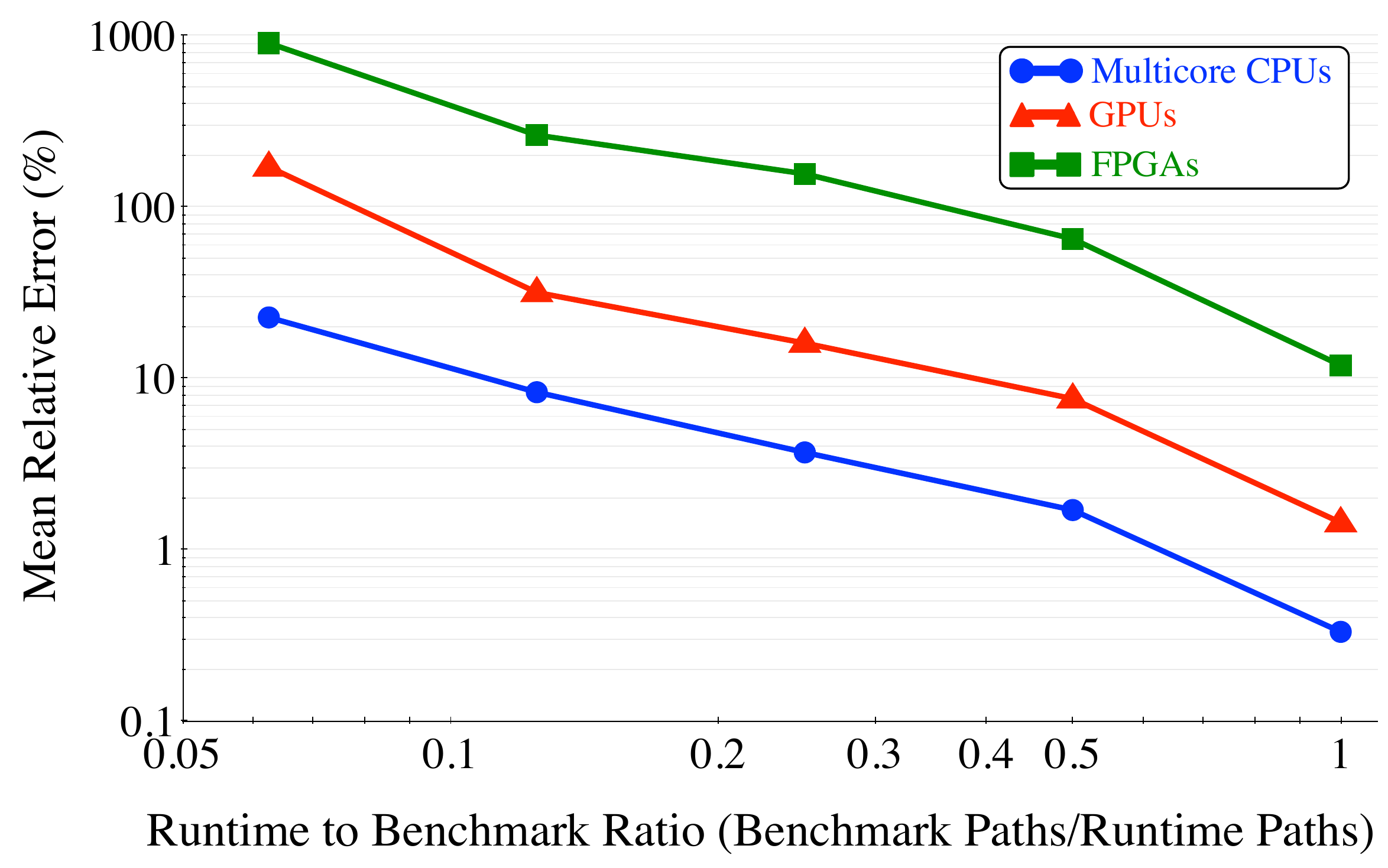}%
        \caption{Device Categories}
        \label{fig:LatencyModelSanityCatogories}
\end{subfigure}\quad%
\begin{subfigure}{.75\columnwidth}
        \includegraphics[width=\columnwidth]{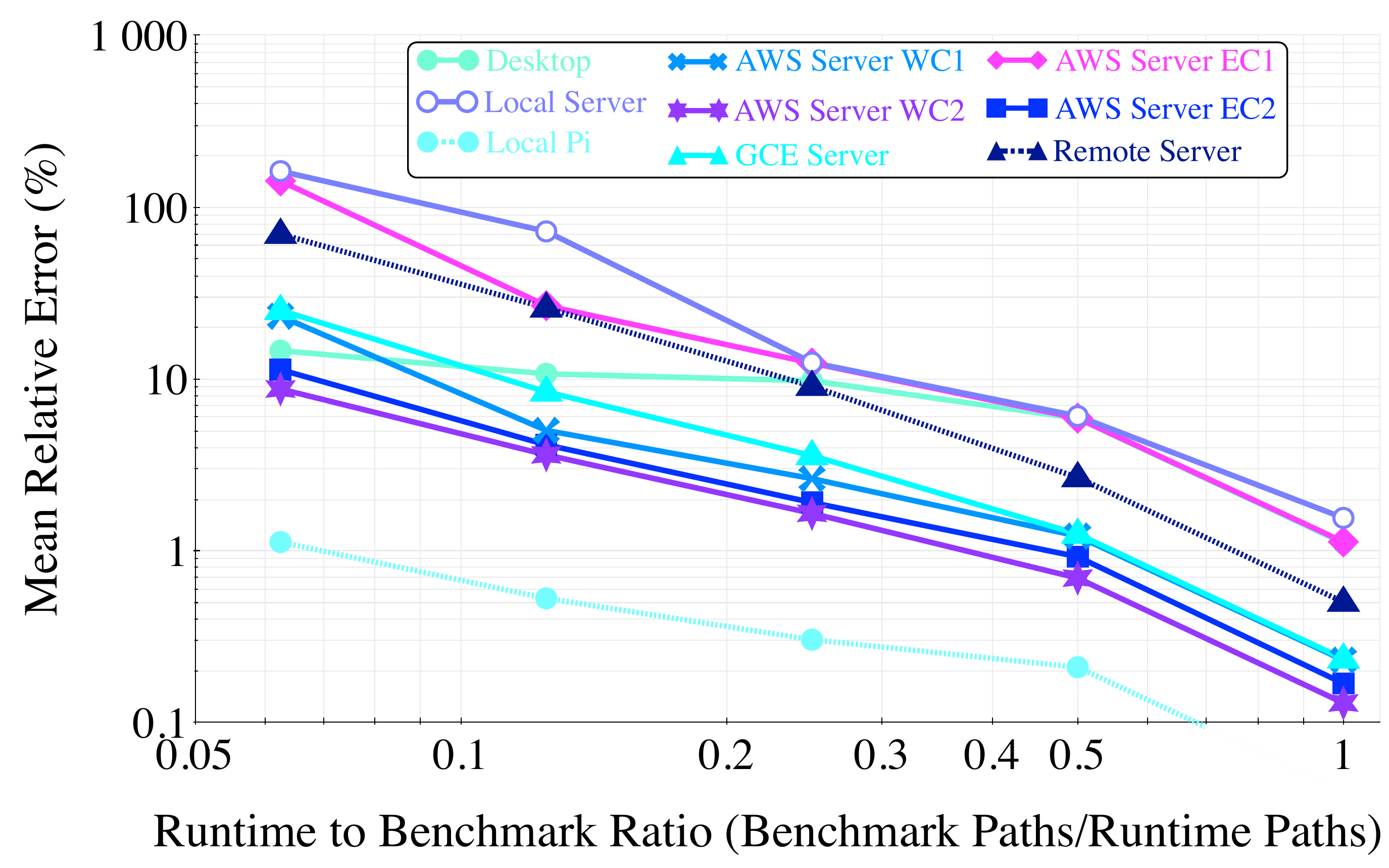}%
        \caption{CPUs}
        \label{fig:LatencyModelSanityCPUs}
\end{subfigure}\vskip\baselineskip%
\begin{subfigure}{.75\columnwidth}
        \includegraphics[width=\columnwidth]{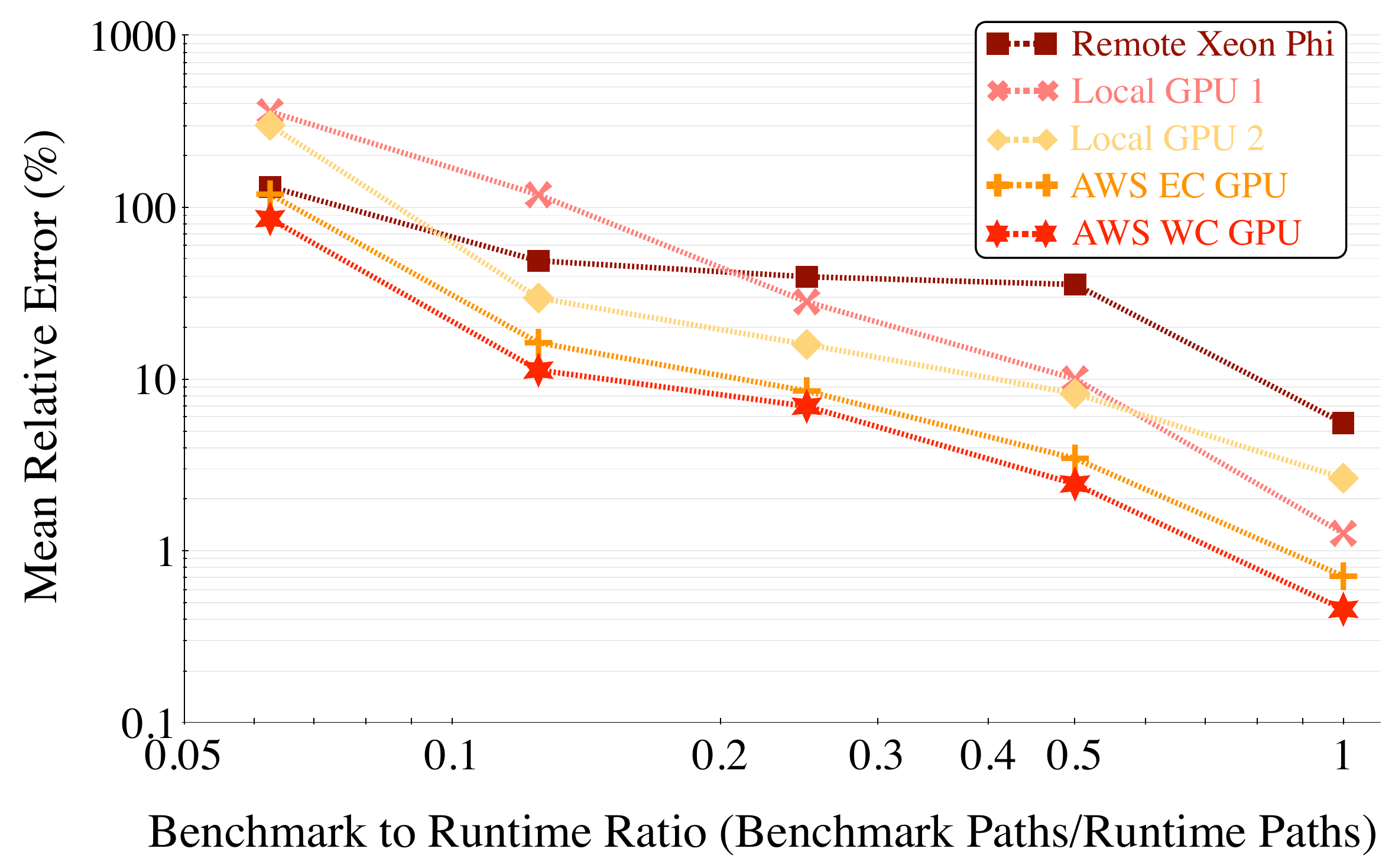}%
        \caption{GPUs}
        \label{fig:LatencyModelSanityGPUs}
\end{subfigure}\quad%
\begin{subfigure}{.75\columnwidth}
        \centering
        \includegraphics[width=\columnwidth]{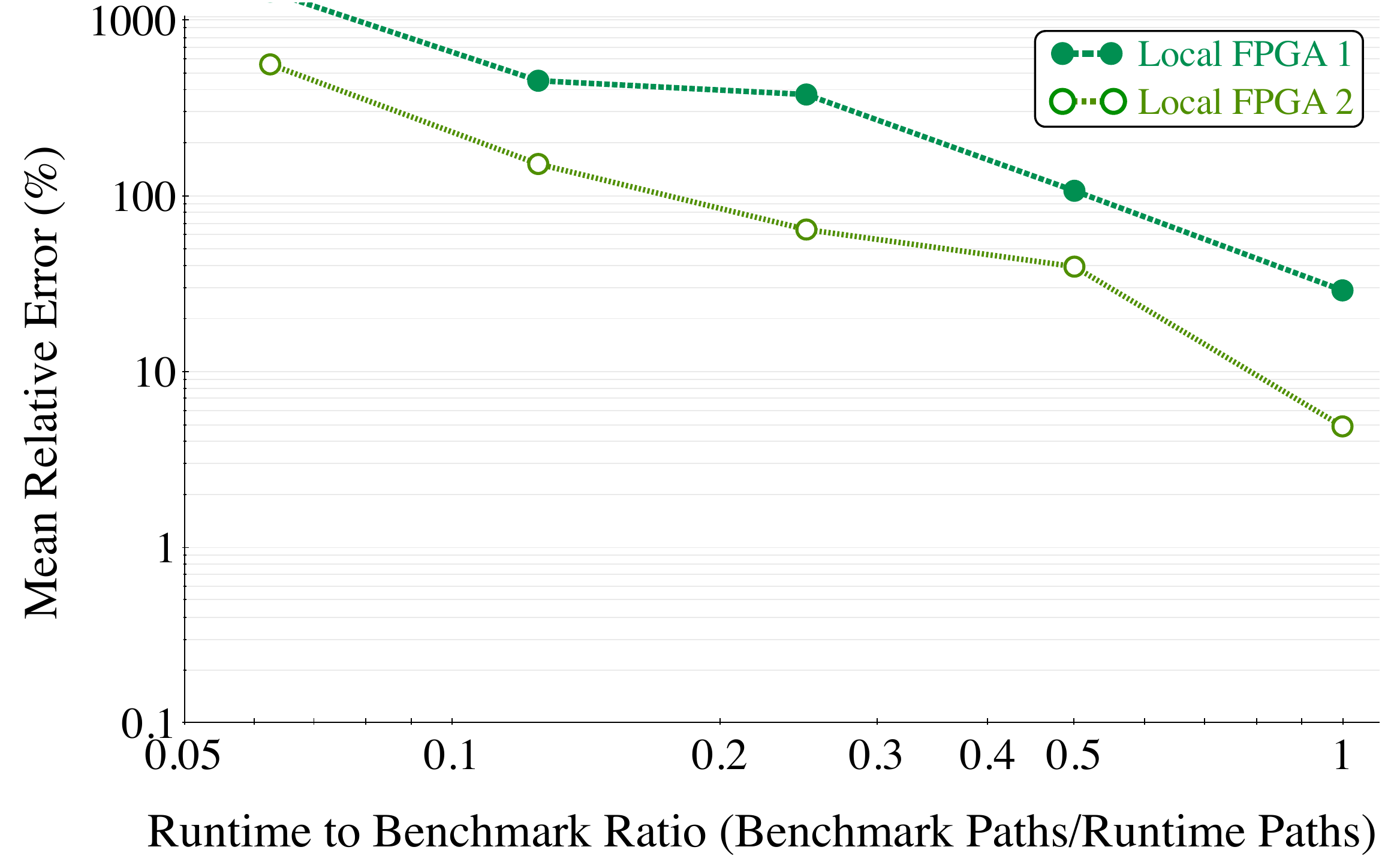}%
        \caption{FPGAs}
        \label{fig:LatencyModelSanityFPGAs}
\end{subfigure}%

\caption{Error of latency models for a total run-time target of 10 minutes ($\frac{4.69s}{\text{task}}$) and varying benchmark time, evaluating model incorporation.}
\label{fig:LatencyModelSanity}
\end{figure*}

\begin{figure*}%
\centering

\begin{subfigure}{.75\columnwidth}
        \includegraphics[width=\columnwidth]{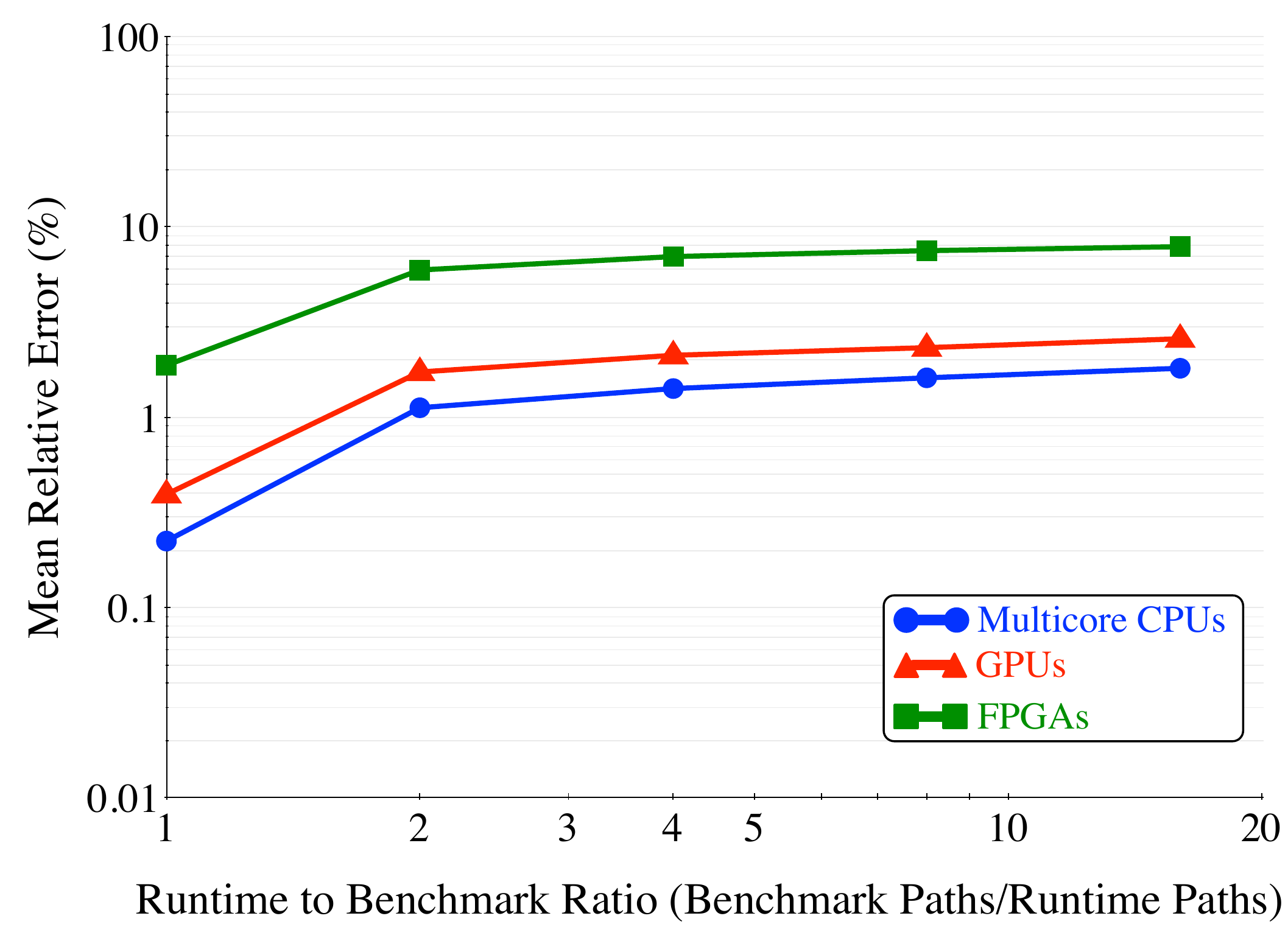}%
        \caption{Device Categories}
        \label{fig:LatencyModelScalingCatogories}
\end{subfigure}\quad%
\begin{subfigure}{.75\columnwidth}
        \includegraphics[width=\columnwidth]{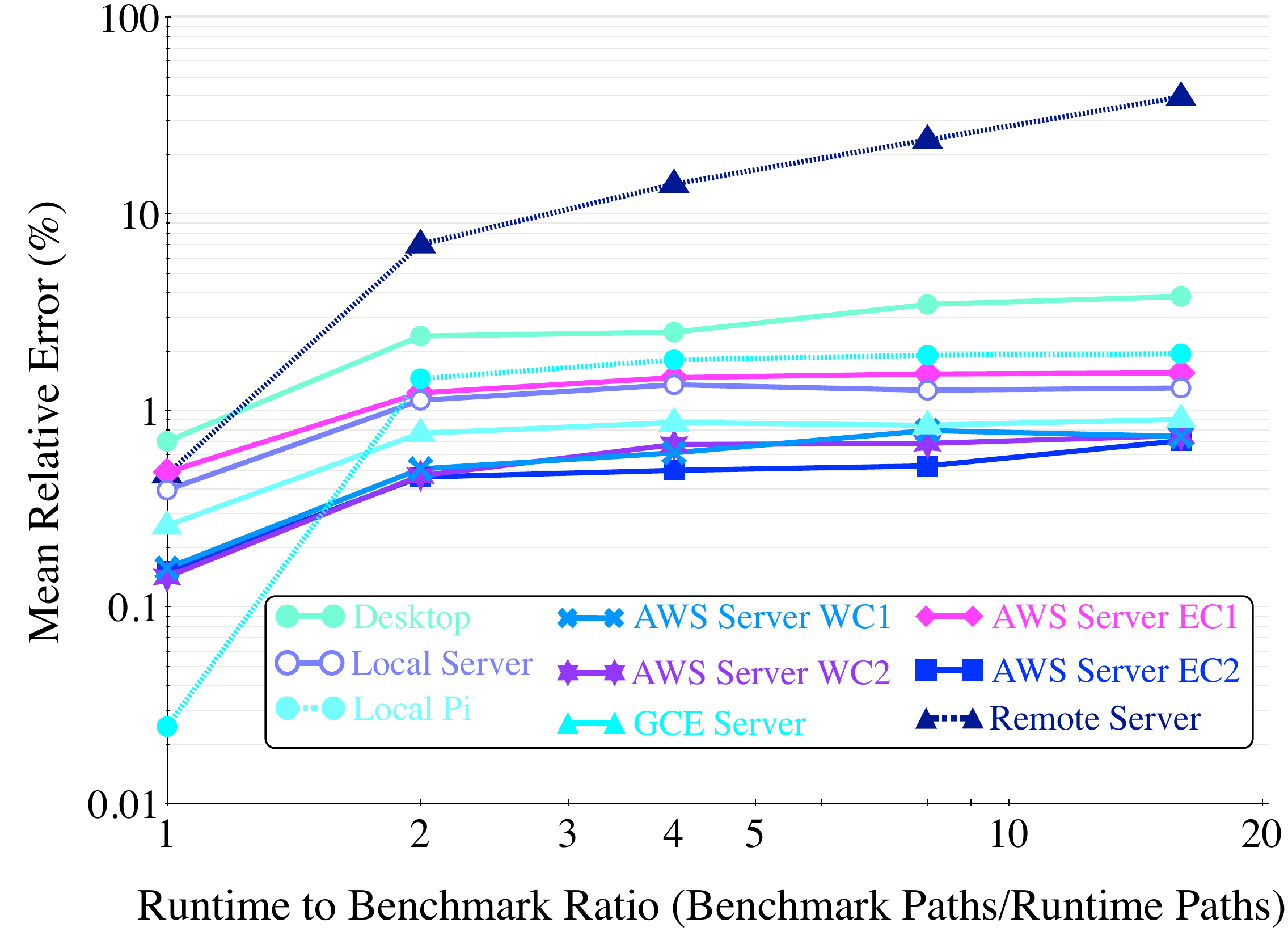}%
        \caption{CPUs}
        \label{fig:LatencyModelScalingCPUs}
\end{subfigure}\vskip\baselineskip%
\begin{subfigure}{.75\columnwidth}
        \includegraphics[width=\columnwidth]{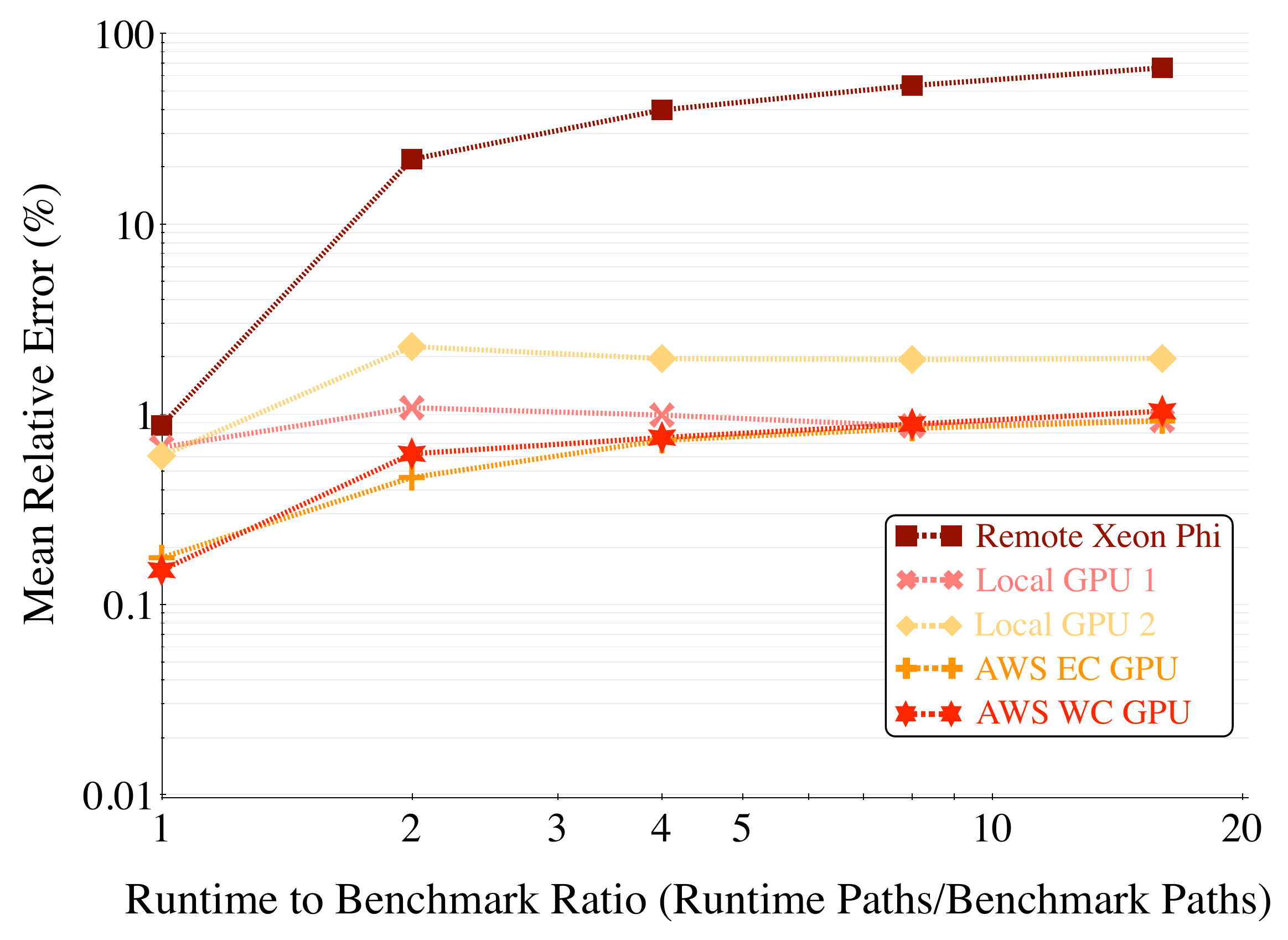}%
        \caption{GPUs}
        \label{fig:LatencyModelScalingGPUs}
\end{subfigure}\quad%
\begin{subfigure}{.75\columnwidth}
        \includegraphics[width=\columnwidth]{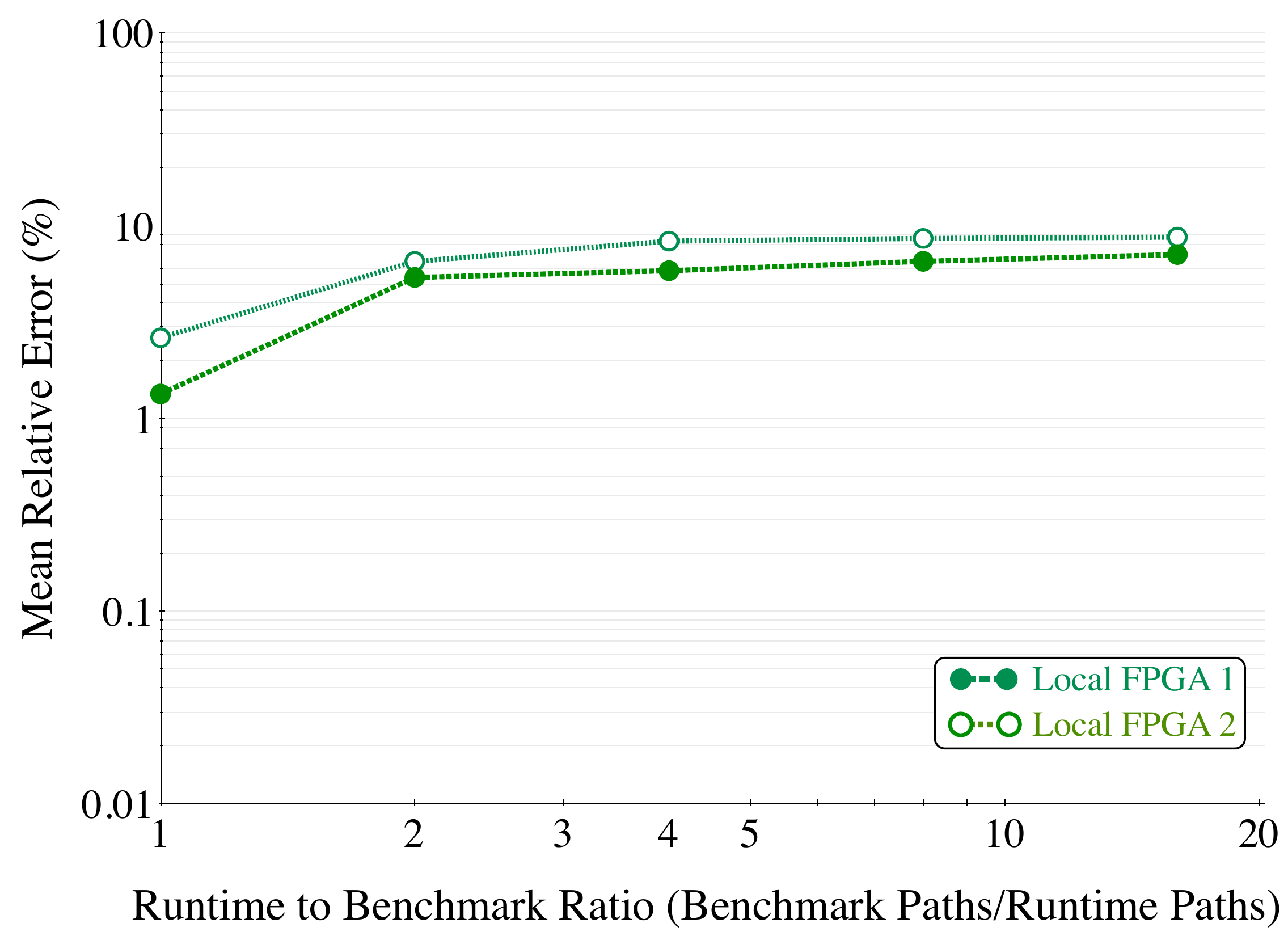}%
        \caption{FPGAs}
        \label{fig:LatencyModelScalingFPGAs}
\end{subfigure}%

\caption{Error of latency models for a fixed benchmark total time of 10 minutes ($\frac{4.69s}{\text{task}}$) and varying run-time targets, evaluating model extrapolation.}
\label{fig:LatencyModelScaling}
\end{figure*}

The latency model results are given in Figures \ref{fig:LatencyModelSanity} and \ref{fig:LatencyModelScaling}. The latency models were evaluated on a per platform basis, as well as the geometric mean of the three platform categories. 

The mean error figures for all of the tasks upon the platform reported, with the error bars being too small to plot. The independent variable is the ratio of the Monte Carlo benchmark vs run-time paths, so as to report on all tasks across each platform.

Figure~\ref{fig:LatencyModelSanity} illustrates that as a longer benchmarking procedure is performed relative to the total fixed run-time of 10 minutes (or 4.69 seconds per task) being predicted by the model, the models became more accurate. Figure~\ref{fig:LatencyModelScaling} shows how the models scale as the run-time prediction target is increased for a fixed benchmarking time of 4.69 seconds per task or 10 minutes in total, and an increasing run-time target. The remote Phi and server models' poor performance are notable data points.


\subsubsection{Accuracy Model}
\begin{figure*}%
\centering

\begin{subfigure}{.45\columnwidth}
        \centering
        \includegraphics[width=\columnwidth]{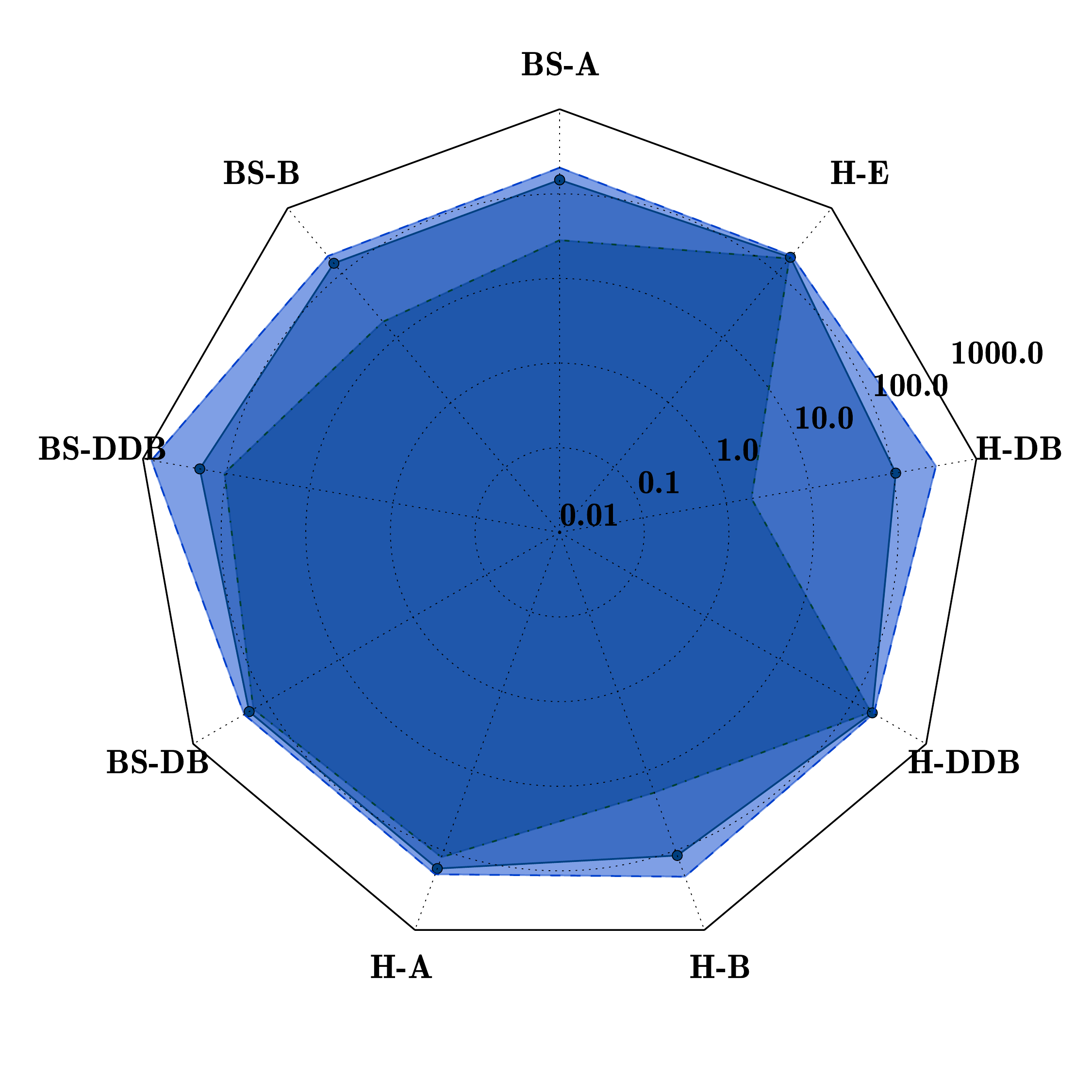}%
        \caption{1:8}
        \label{fig:AccuracyModelSanity0125}
\end{subfigure}\hfill%
\begin{subfigure}{.45\columnwidth}
        \centering
        \includegraphics[width=\columnwidth]{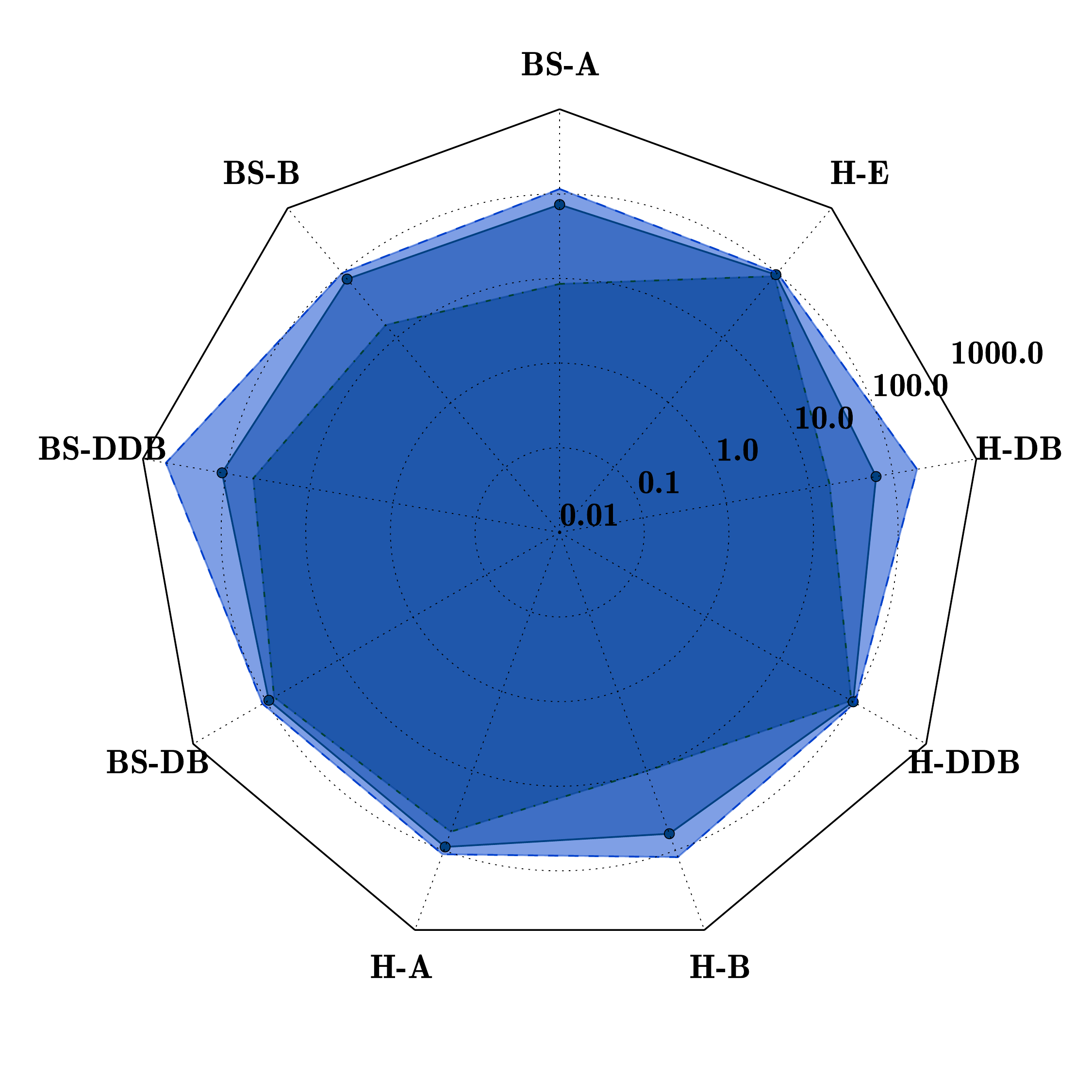}%
        \caption{1:4}
        \label{fig:AccuracyModelSanity025}
\end{subfigure}\hfill
\begin{subfigure}{.45\columnwidth}
        \centering
        \includegraphics[width=\columnwidth]{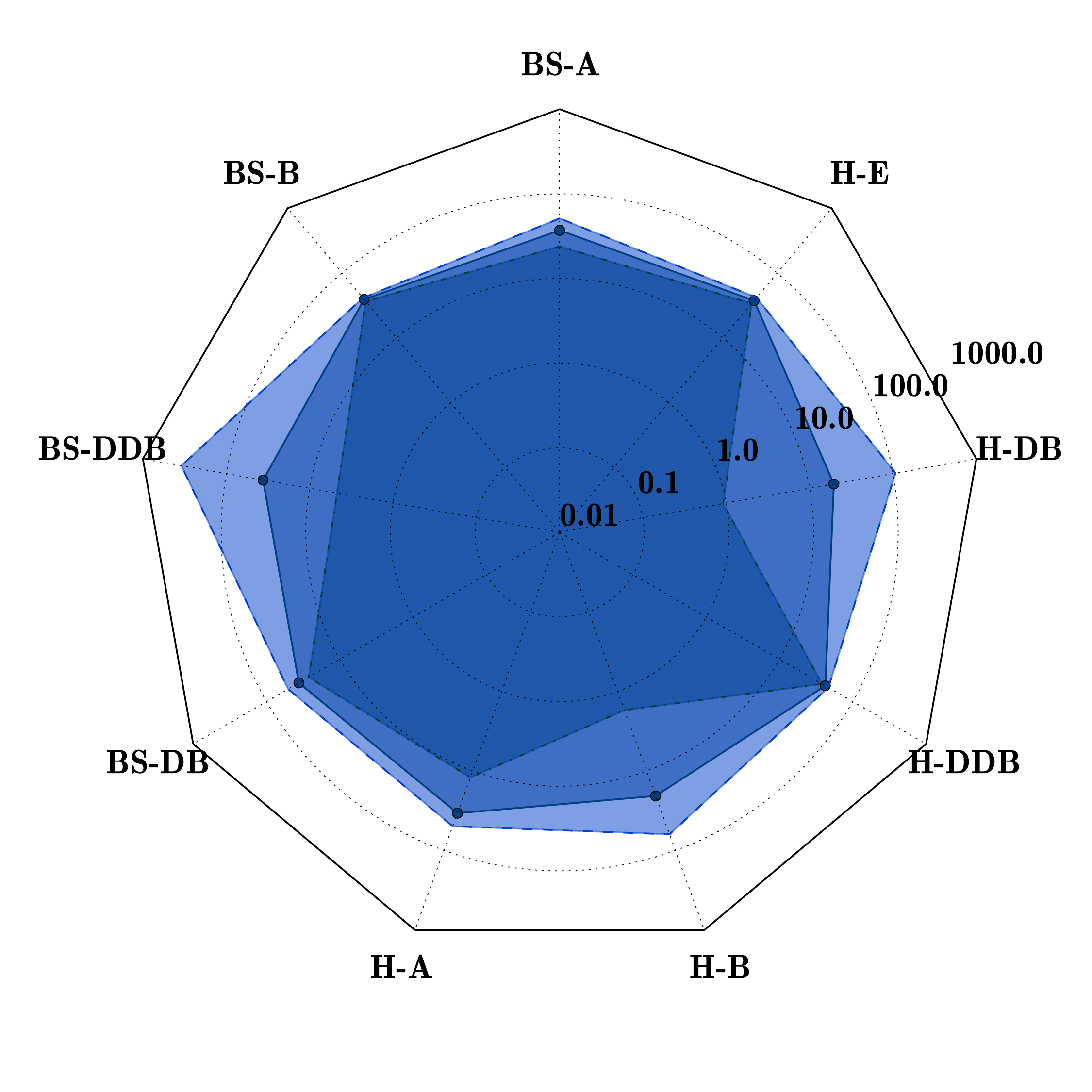}%
        \caption{1:2}
        \label{fig:AccuracyModelSanity05}
\end{subfigure}\hfill
\begin{subfigure}{.45\columnwidth}
        \centering
        \includegraphics[width=\columnwidth]{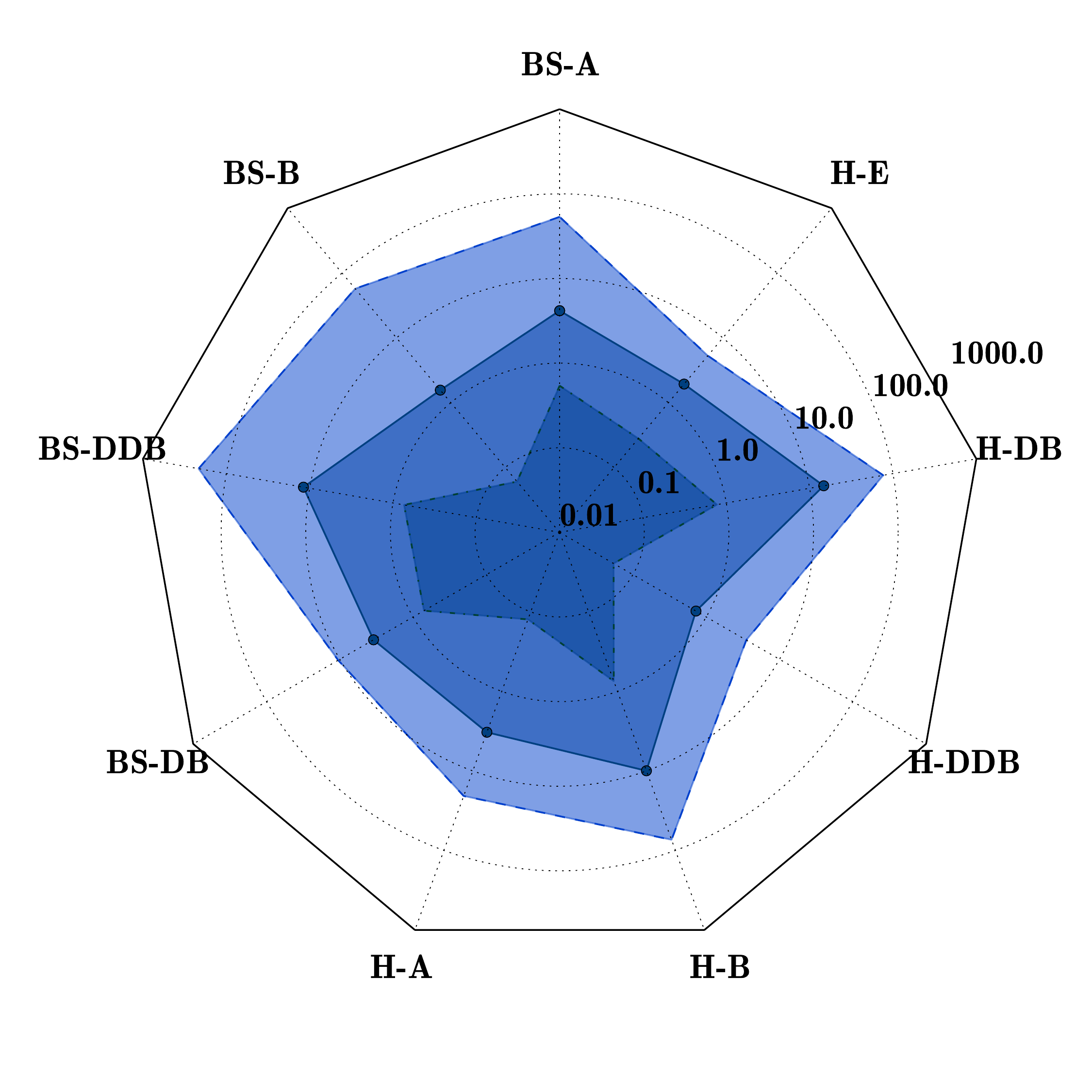}%
        \caption{1:1}
        \label{fig:AccuracyModelSanity10}
\end{subfigure}%

\caption{Percent error of accuracy models for a fixed run-time target and varying benchmark time, evaluating model incorporation. Ratio is expressed as \emph{Benchmark Paths}~:~\emph{Run-time Paths}. The innermost region represents the minimum error, the middle region the geometric mean relative error and the outermost the maximum.}

\label{fig:AccuracyModelSanity}
\end{figure*}

\begin{figure*}%
\centering

\begin{subfigure}{.45\columnwidth}
        \includegraphics[width=\columnwidth]{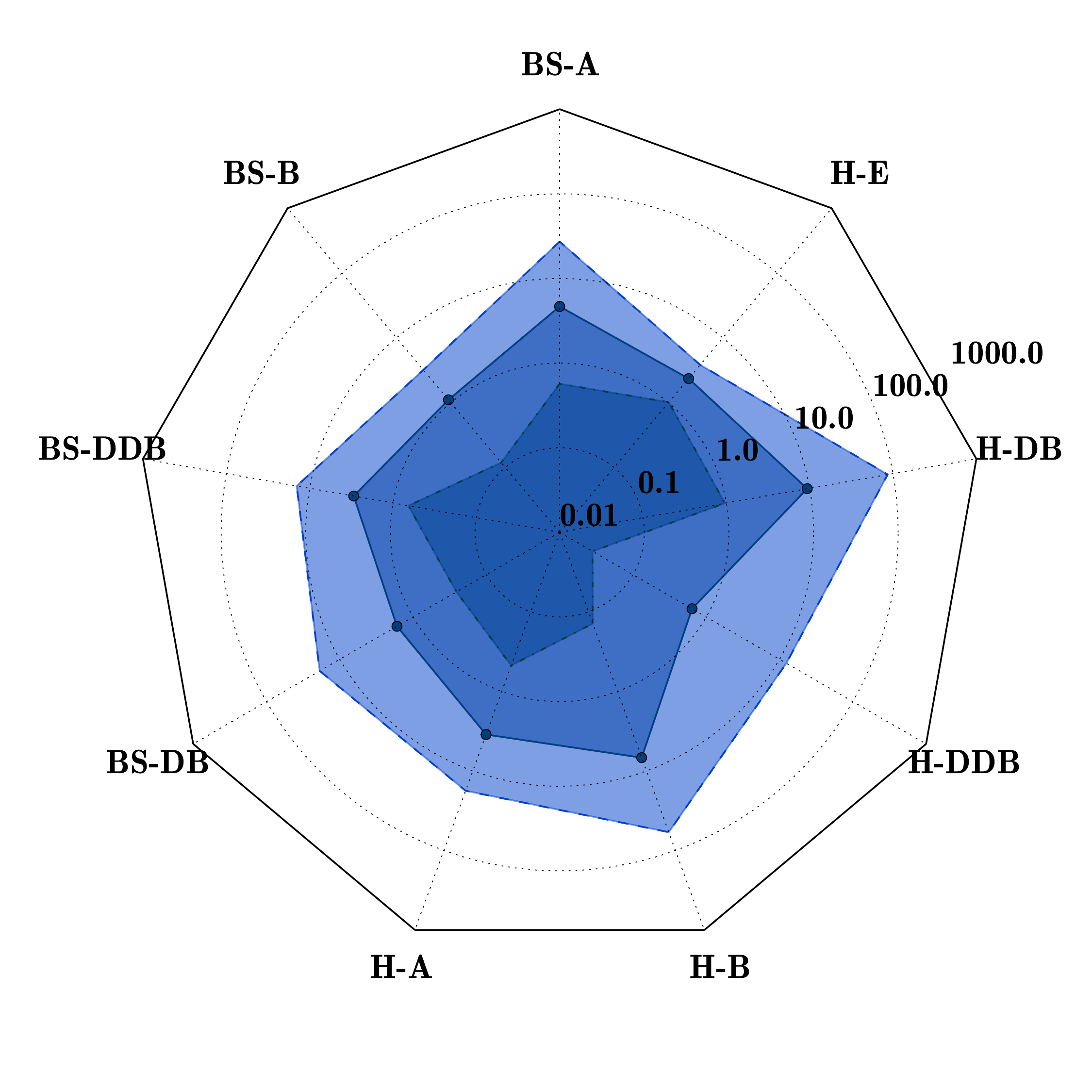}%
        \caption{1:2}
        \label{fig:AccuracyModelSanity20}
\end{subfigure}\hfill%
\begin{subfigure}{.45\columnwidth}
        \includegraphics[width=\columnwidth]{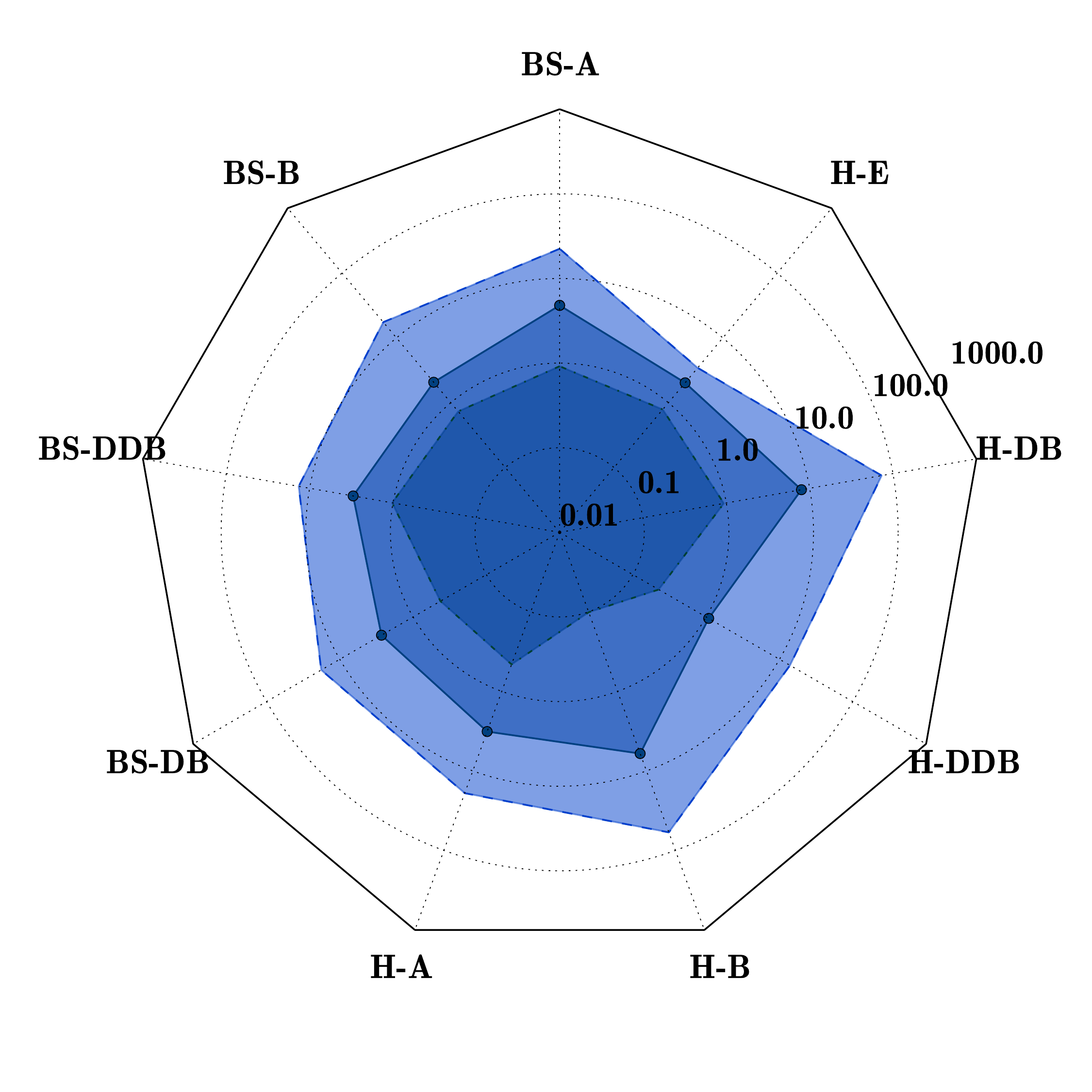}%
        \caption{1:4}
        \label{fig:AccuracyModelSanity40}
\end{subfigure}\hfill
\begin{subfigure}{.45\columnwidth}
        \includegraphics[width=\columnwidth]{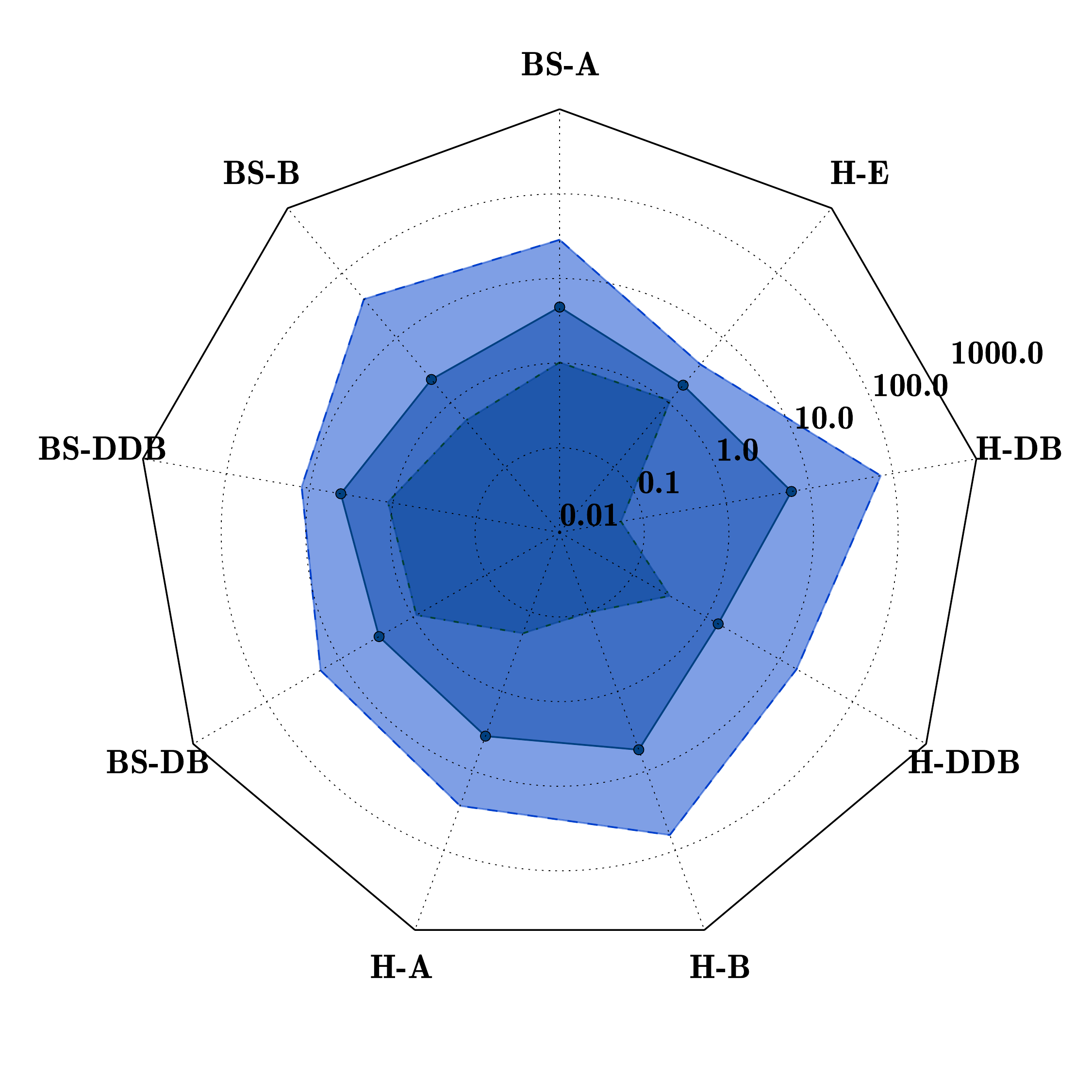}%
        \caption{1:8}
        \label{fig:AccuracyModelSanity50}
\end{subfigure}\hfill%
\begin{subfigure}{.55\columnwidth}
        \includegraphics[width=.8\columnwidth]{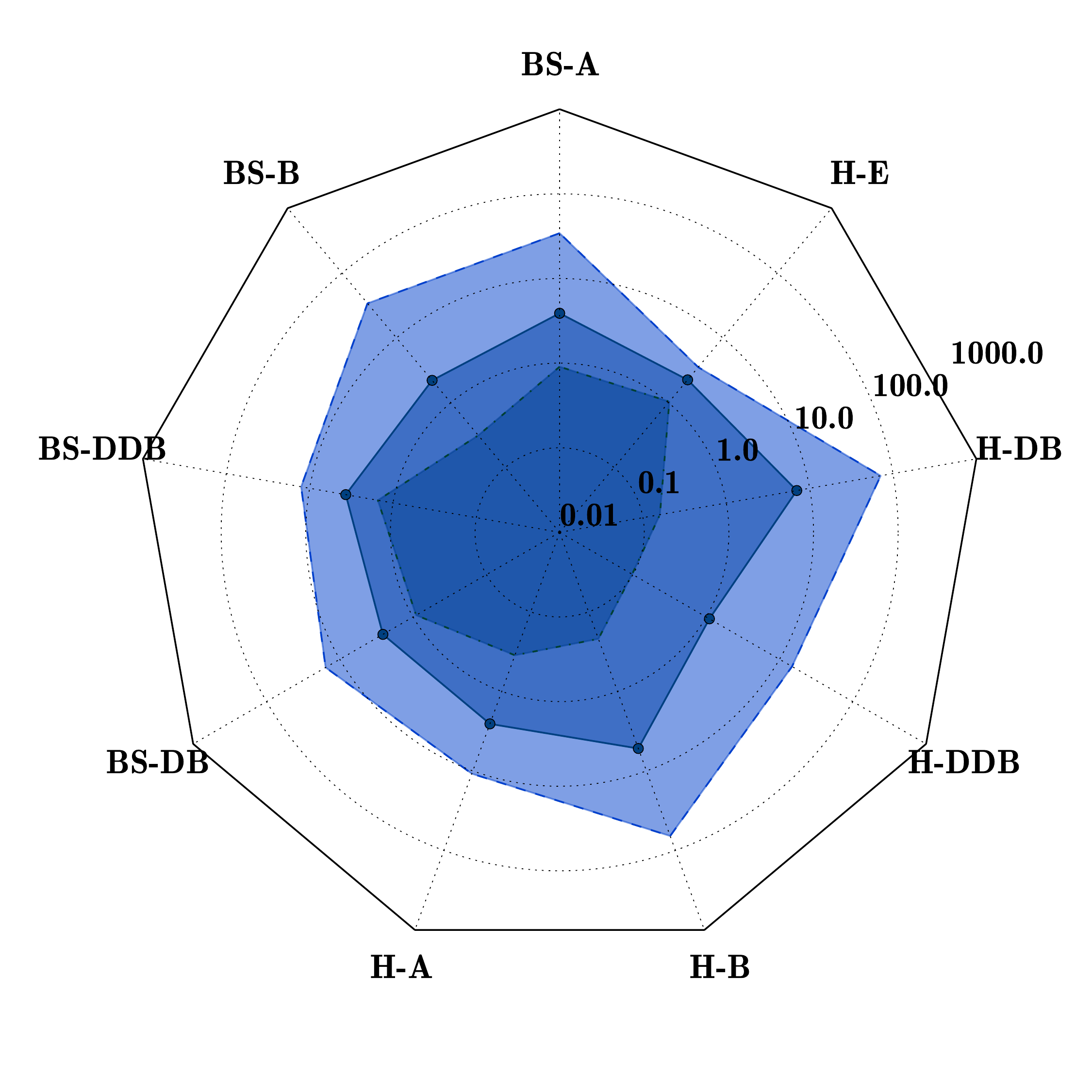}%
        \caption{1:16}
        \label{fig:AccuracyModelSanity160}
\end{subfigure}%

\caption{Percent error of accuracy models for a fixed benchmark time and varying run-time, evaluating model extrapolation. Ratios and the ordering of the regions are the same as in Figure \ref{fig:AccuracyModelSanity}.}

\label{fig:AccuracyModelScaling}
\end{figure*}

The accuracy model results are given in Figures \ref{fig:AccuracyModelSanity} and \ref{fig:AccuracyModelScaling}. The accuracy model results are presented as  minimum, geometric mean and maximum of the model results within the pricing task categories. Similar to the latency model, the ratio of benchmark to run-time paths is the independent variable.


Figure \ref{fig:AccuracyModelSanity} shows how the accuracy models become increasingly predictive with additional benchmarking. For all of the task categories,  the minimum, mean and maximum errors all decrease as more benchmarking is performed. Figure \ref{fig:AccuracyModelScaling} shows how the models remain stable as the run-time target is increased. Hence, similar to the latency model results, the models scale well for more than an order of magnitude.

\subsection{Discussion}

The incorporation property of the models is demonstrated in Figures \ref{fig:LatencyModelSanity} and \ref{fig:AccuracyModelSanity}. This means that the predictive capability of the models improves as additional information is provided to the models. 

As Figures \ref{fig:LatencyModelScaling} and \ref{fig:AccuracyModelScaling} illustrate, the models also have the extrapolation property. There is a relatively minor increase in latency and accuracy error for run-times considerably longer than the benchmarking time.

The relatively poor latency model performance of the Remote Phi and Server platforms is explained by the benchmarking time being too short to accurately solve for the true coefficient and constant values. This is due to long network round-trip time that both platforms experienced, where the almost all of the benchmarking time is spent on communication.

The tasks with Heston underlyings present a relatively high maximum accuracy error, between 10\% and 100\%. However, as can be seen by the task category geometric mean these error average out to a considerably lower figure, allowing for these models to still be useful for modelling groups of tasks.



\section{Domain Allocation Approach Evaluation}
\label{sec:DomainPartitionerEvaluation}
In this section we describe our evaluation of the allocation approaches that make use of domain knowledge provided through the metric models, machine learning and MILP. We first characterise the performance of the domain allocation approaches with respect to problem size and problem non-linearity using synthetic data. We then verify this characterisation by applying the different allocation approaches to the tasks and platforms described in Tables \ref{table:TaskDetails} and \ref{table:PlatformOverview}.

We report on the time required by the domain allocation approach algorithms as well as the quality of the solution returned with respect to the proportional allocation heuristic. For an allocation approach to be acceptable, it needs to cope with a wide variety of allocation problems in we what we heuristically define as a reasonable amount of time, 10 minutes, while providing a significant improvement over the allocation returned by the heuristic approach.



\subsection{Experimental Setup}

\subsubsection{Synthetic Data Generation Procedure}
Drawing upon Braun et al's work~\cite{Braun11Heuristics}, we have used the following procedure ($s(\tau,\mu,\theta_\tau,\theta_\mu,\omega_\tau,\omega_\mu,\psi)$) to generate the synthetic co-efficient ($\boldsymbol{\delta}$) and constant ($\boldsymbol{\gamma}$) matrices so to evaluate the different approaches to allocation:

(1) Construct the baseline vector ($\vec{x}$) and initial matrix ($\boldsymbol{Y}$). $\vec{x}$ is $\tau$ uniformly distributed integer elements, bounded by the task heterogeneity factor ($\theta_\tau$). $\boldsymbol{Y}$, is $\mu \times \tau$ uniformly distributed integer elements, bounded by the platform heterogeneity factor ($\theta_\mu$):
$$ x_j \in [1,\theta_\tau] \quad j = \{1,2,\ldots,\tau\}, $$
$$Y_{i,j} \in [1,\theta_\mu] \quad i = \{1,2,\ldots,\mu\},j = \{1,2,\ldots,\tau\}.$$

(2) Construct the $\boldsymbol{\delta}$ matrix by multiplying the elements of each row of $\boldsymbol{Y}$ and of $\vec{x}$. i.e. $$\delta_{i,j} = x_j Y_{i,j}  \quad i = \{1,2,\ldots,\mu\},j = \{1,2,\ldots,\tau\}.$$

(3) Sort the first $\tau\omega_\tau$ columns of the $\boldsymbol{\delta}$ matrix, and the first $\mu\omega_\mu$ rows, where $\omega_\tau$ and $\omega_\mu$ are the degree of task and platform consistency.

(4) Construct the $\boldsymbol{\gamma}$ matrix by repeating steps 1-3, however then multiply the resulting matrix by the task constant to coefficient ratio ($\psi$), i.e. the $\gamma$ to $\beta$ ratio in the latency metric model.

\subsubsection{Procedure Parameter Values}
The parameters varied in our experiments, in conjunction with the procedure above are provided in Table \ref{table:SyntheticDetails}. The four cases consider a range of different scenarios, from completely homogeneous, consistent to extremely heterogeneous and inconsistent platforms and tasks, using the values from Braun et al's study~\cite{Braun11Heuristics}.

\begin{table}
\begin{center}
\caption{Synthetic task-platform generation parameters. Columns are platform and task heterogeneity ($\theta_\mu,\theta_\tau$), and platform and task consistency ($\omega_\mu,\omega_\tau$).}
\label{table:SyntheticDetails}
\centering
\begin{tabular}{ccccc}
\parbox[c]{1.5cm}{\center Case Designation} & $\theta_\mu$ & $\omega_\mu$ & $\theta_\tau$ & $\omega_\tau$\\ \hline\\
    Hom-Con & 10 & 1.0 & 100 & 1.0 \\
    Het-Con & 100 & 1.0 & 3000 & 1.0 \\
    Het-Mix & 100 & 0.5 & 3000 & 0.5 \\
    Het-Inc & 100 & 0.0 & 3000 & 0.0
\end{tabular}
\end{center}
\end{table}

\subsection{Allocation Characterisation Results}
\subsubsection{Synthetic Data Characterisation}
\begin{figure*}
\centering
\caption{Characterisation of allocation approaches using synthetic data.}


\begin{subfigure}{\columnwidth}
        \includegraphics[width=.9\columnwidth]{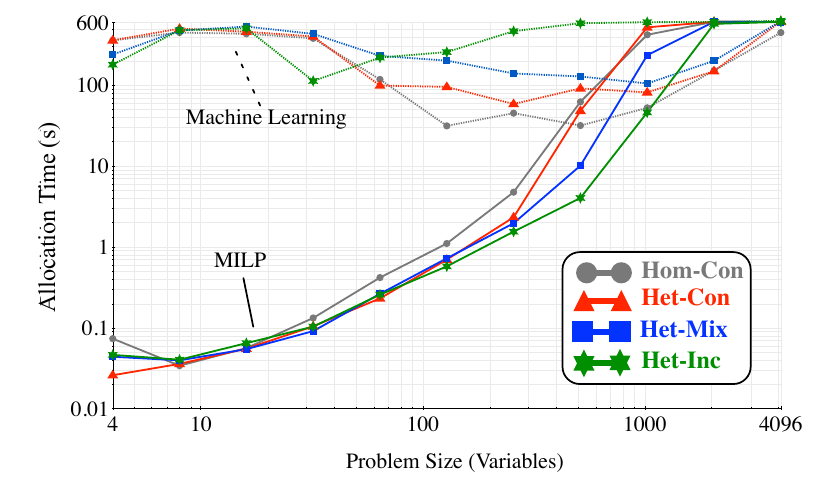}%
        \caption{Impact of allocation problem size on allocation time. The constant to coefficient ratio is 1}
        \label{fig:SyntheticPartitionerVariablesLatency}
\end{subfigure}\hfill%
\begin{subfigure}{\columnwidth}
        \includegraphics[width=.9\columnwidth]{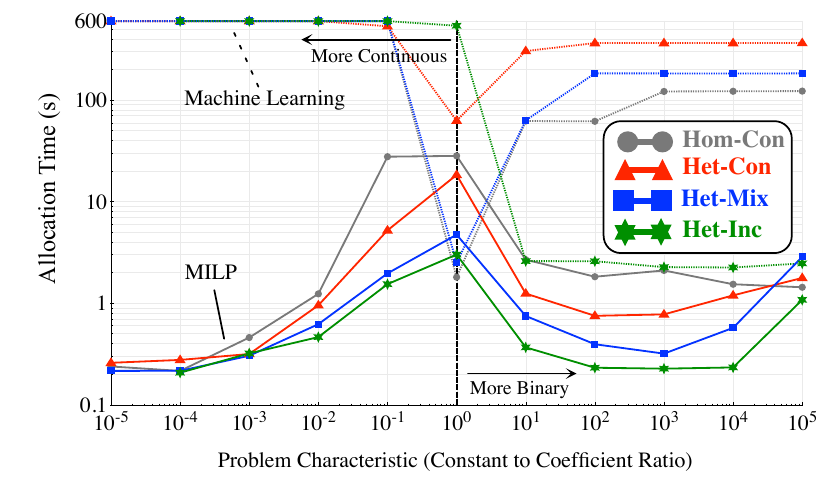}%
        \caption{Impact of problem characteristic on allocation time. The number of variables is 1024}
        \label{fig:SyntheticPartitionerRatioLatency}
\end{subfigure}\vskip\baselineskip%
\begin{subfigure}{\columnwidth}
        \includegraphics[width=.9\columnwidth]{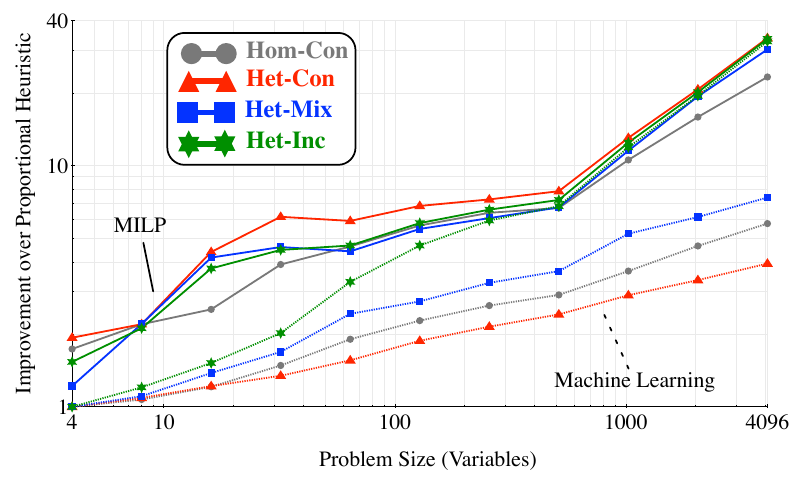}%
        \caption{Improvement for varied problem sizes. The constant to coefficient ratio is 1.}
        \label{fig:SyntheticPartitionerVariablesImrpovement}
\end{subfigure}\hfill%
\begin{subfigure}{\columnwidth}
        \centering
        \includegraphics[width=.9\columnwidth]{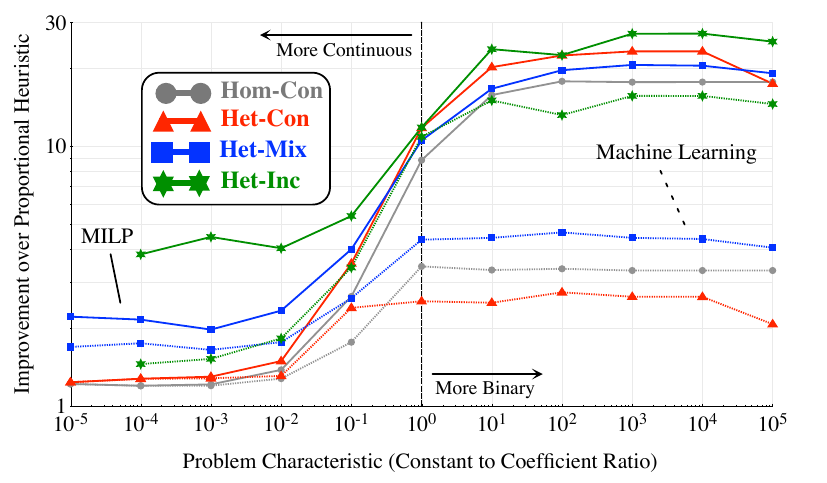}%
        \caption{Improvement for varied problem characteristics. The number of variables is 1024}
        \label{fig:SyntheticPartitionerRatiosImprovement}
\end{subfigure}%

\label{fig:SyntheticCharacterisation}
\end{figure*}

The results of the allocation characterisation can be seen in Figure \ref{fig:SyntheticCharacterisation}. For the allocation times (Figures \ref{fig:SyntheticPartitionerVariablesLatency} and \ref{fig:SyntheticPartitionerRatioLatency}), a timeout of 600 seconds (or 10 minutes) was set, the same time given to the benchmarking described in the previous section.

Similarly for the quality of the solution relative to the proportional heuristic (Figures \ref{fig:SyntheticPartitionerVariablesImrpovement} and \ref{fig:SyntheticPartitionerRatiosImprovement}), we found that both the MILP and machine learning task allocations' improvements over the heuristic are a function of problem variables and constant to coefficient ratio. 

\begin{figure*}
\centering
\caption{Allocation approaches using heterogeneous platforms from Table \ref{table:PlatformOverview} and derivatives pricing tasks from Table \ref{table:TaskDetails}. Smaller latency and accuracy values are better.}
\includegraphics[width=0.9\textwidth]{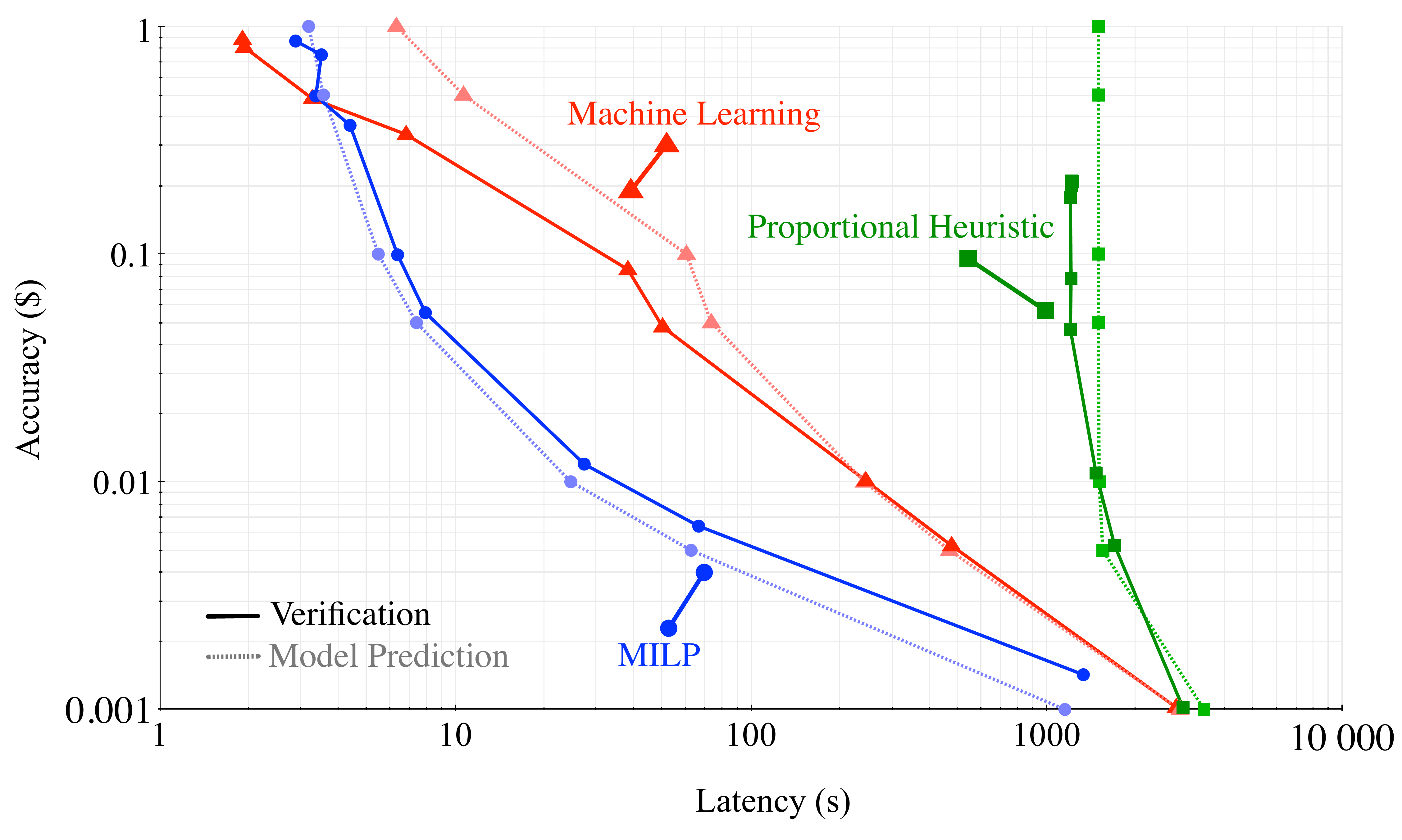}
\label{figure:ModelCurves}
\end{figure*}

\subsubsection{Practical Verification}
While the characterisation of the allocation approaches using synthetic data provides insight, we have verified these these results with real platform and task data in Figure \ref{figure:ModelCurves}. We put the portfolio of pricing tasks in Table \ref{table:TaskDetails} through the allocation approaches for the platforms in Table \ref{table:PlatformOverview} over a range of accuracies. We then ran the generated task allocations, and measured the domain metrics of latency and accuracy.


\subsection{Discussion}

Broadly, the machine learning-based task allocation approach was soon limited by the timeout, as evidenced by Figure \ref{fig:SyntheticPartitionerVariablesLatency}, while the MILP task allocation's time grows exponentially as a function of the number of variables. As the ratio between the coefficient ($\beta$) and constant ($\gamma$) component is varied in Figure \ref{fig:SyntheticPartitionerRatioLatency}, there is a peak latency centred around 1, reflecting the considerable linear and non-linear allocation problems that both have to be solved and balanced. The machine learning approaches perform well at this inflection point, while the MILP approach is at its relative worst.

In terms of improvement over the heuristic allocation, Figures \ref{fig:SyntheticPartitionerVariablesImrpovement} and \ref{fig:SyntheticPartitionerRatiosImprovement}, the trends can be explained by the potential for improvement. The linear trend with respect to problem size is due to the increased potential for improvement that larger problems allow. Similarly, for the constant to coefficient ratio, as the constant becomes more dominant, there is increased scope for improvement as the heuristic is further from its optimal condition.


Figure \ref{figure:ModelCurves} illustrate that the allocation approaches using the metric models are close to what is measured in reality. The differences between the predictions and run-time measurements are well within the error of the metric models. Furthermore, both the domain knowledge-based machine learning and MILP-based allocation approaches are orders of magnitude more efficient than that suggested by the proportional heuristic for problems with strong non-linear characteristics, i.e. derivatives pricing tasks with 95\% confidence intervals greater than \$0.005.

\section{Conclusion}
\label{sec:Conclustion}

In this paper we have described and demonstrated in practice that a domain specific approach to heterogeneous computing offers two features beyond portable execution.

Firstly, using image filtering, linear algebra arithmetic and derivatives pricing as example domains, we described how domain metric models derived from the application domain provide a natural means to characterise a task on a heterogeneous platform. 

We evaluated the metric models of latency and accuracy for the derivative pricing domain practically. We found that when using an online benchmarking procedure, these domain metric models incorporate additional information to improve predictions, and extrapolate well as tasks increase in scale. 

\begin{figure}
\centering
\includegraphics[width=.55\textwidth]{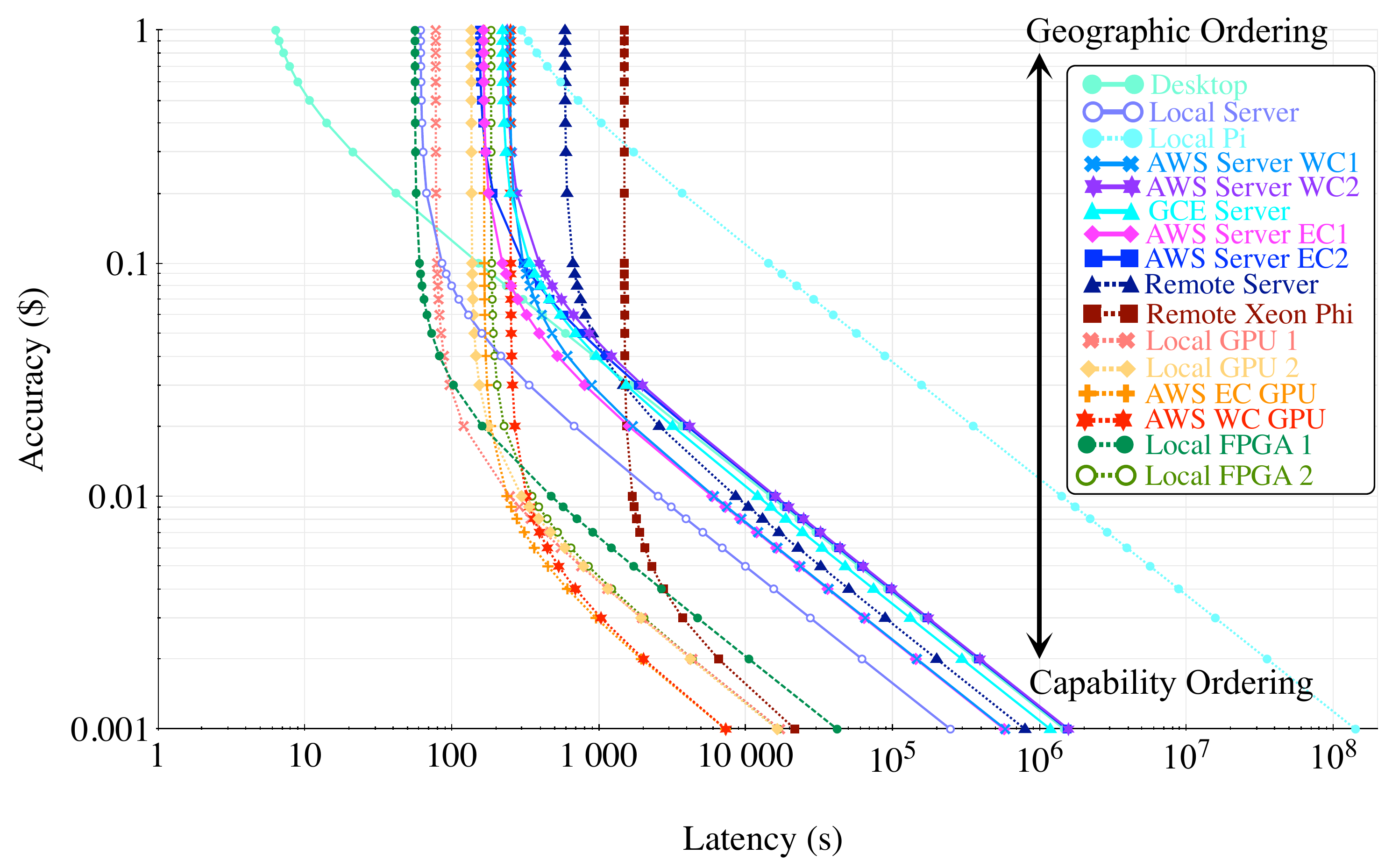}
\caption{Heterogeneous task-platform metric curves. Smaller accuracy and latency values are better.}
\label{fig:PlatformCurves}
\end{figure}

These metric models are an accessible way to visualise the design space of the available heterogeneous platforms for domain programmers, such as Julia, as illustrated in Figure \ref{fig:PlatformCurves}. As is to be expected, when the accuracy requirement is low, constant communication time dominate and the platform makespans are geographically ordered, but when high accuracy is required, the compute dominates and the makespans order according to the measured computational capability of the platforms. 

Secondly, we described how the metric models for multiple platforms can be combined into a constrained optimisation program. We also showed domain specific knowledge can allow this allocation to be relaxed, transforming the binary problem to a more tractable, mixed integer form.

\begin{figure}
\centering
\includegraphics[width=.55\textwidth]{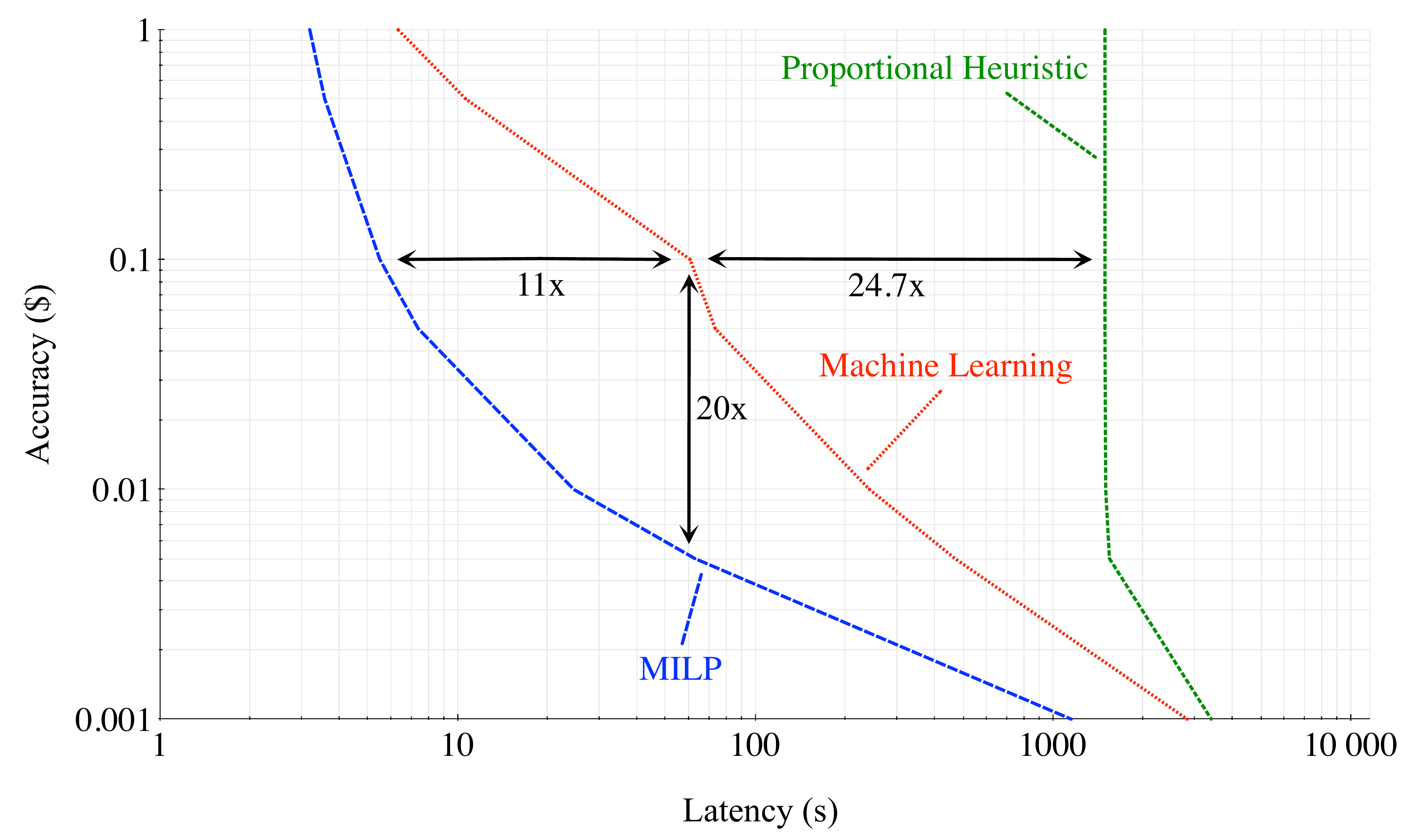}
\caption{Differing domain specific task allocation approaches. Smaller accuracy and latency values are better.}
\label{fig:ParetoCurvesModelsVerification}
\end{figure}

We described and evaluated three approaches for solving this allocation problem, heuristic, machine learning and MILP. Our evaluation, making use of both synthetic data as well as the derivatives pricing examples, demonstrated that both MILP and machine learning can produce viable task allocation in a practical amount of time whilst outperforming the heuristic approach by up to two orders of magnitude, as illustrated in Figure \ref{fig:ParetoCurvesModelsVerification}.

Beyond the practical benefits, our domain specific methodology makes distributed, heterogeneous computing platforms \emph{accessible} to domain users, such as Julia in the Introduction. Our approach shows how to abstract away details of implementation into choices about the nature of computational results. 
We only require that Julia balances her objectives to use heterogeneous computing effectively.

\subsubsection*{Future Work}
Future directions for this work include increasing both the allocation problem sizes as well as the number of platforms utilised. A further direction is in increasing the degree of heterogeneity, both in terms of the problems considered as well as more varied computing resources.

Another area of ongoing work is optimisation of the MILP software used. Improvements being considered are seeding the optimiser with the proportional heuristics proposed in this paper, as well as ordering the heuristics applied within the optimiser more carefully.

\ifCLASSOPTIONcompsoc
  \section*{Acknowledgments}
\else
  \section*{Acknowledgment}
\fi
Conversations with Prof George Constantinides, Dr Victor Magron, Dr Eric Kerrigan, Dr Andrea Suardi, Mr Shane Fleming and Mr Andrea Picciau have been invaluable.

We are grateful for the funding support from the South African National Research Foundation and Oppenheimer Memorial Trust. We have also benefited from the support of the Nallatech, Altera, Xilinx, Intel and Maxeler university programs.

\ifCLASSOPTIONcaptionsoff
  \newpage
\fi



\bibliographystyle{IEEEtran}
\bibliography{IEEEabrv,TOPDS2015.bib}
%



%

\vspace{-15 mm}

\begin{IEEEbiographynophoto}{Gordon Inggs}
received the B.Sc.(Eng.) \emph{Hons}. and M.Sc.(Eng.) degrees in electrical engineering from the University of Cape Town, and recently completed his Ph.D. degree in Computer Engineering at Imperial College London. His research interests include distributed, heterogeneous computing, and domain specific programming tools.
\end{IEEEbiographynophoto}

\vspace{-15 mm}

\begin{IEEEbiographynophoto}{David B. Thomas} 
received the M.Eng. and Ph.D. degrees in computer science from Imperial College London. He is a Lecturer with the Electrical and Electronic Engineering Department, Imperial College London. His research interests include hardware-accelerated cluster computing, Monte Carlo simulation, random number generation, and financial computing.
\end{IEEEbiographynophoto}

\vspace{-15 mm}

\begin{IEEEbiographynophoto}{Wayne Luk} 
received the M.A., M.Sc., and D.Phil. degrees in engineering and computing science from the University of Oxford. He is a Professor of computer engineering with Imperial College London, and a Visiting Professor with Stanford University and Queens University Belfast. His research interests include theory and practice of customising computing for application domains, such as multimedia, networking, and finance.
\end{IEEEbiographynophoto}





\end{document}